\documentclass[pdflatex,sn-mathphys,Numbered]{sn-jnl}

\usepackage{graphicx}
\usepackage{multirow}
\usepackage{amsmath,amssymb,amsfonts}
\usepackage{amsthm}
\usepackage{mathrsfs}
\usepackage{xcolor}
\usepackage{textcomp}
\usepackage{manyfoot}
\usepackage{booktabs}
\usepackage{algorithm}
\usepackage{algorithmicx}
\usepackage{algpseudocode}
\usepackage{listings}

\usepackage{colortbl}
\usepackage{pifont}
\usepackage{enumitem}
\usepackage{tikz}
\usetikzlibrary{arrows.meta, positioning, calc, shapes.geometric, fit, backgrounds, patterns}
\usepackage{pgfplots}
\pgfplotsset{compat=1.16}

\theoremstyle{thmstylethree}\newtheorem{definition}{Definition}\theoremstyle{thmstyletwo}\newtheorem{remark}{Remark}

\usepackage{letltxmacro}
\makeatletter
\LetLtxMacro{\oldincludegraphics}{\includegraphics}
\renewcommand{\includegraphics}[2][]{\IfFileExists{#2}{\oldincludegraphics[#1]{#2}}{\IfFileExists{#2.pdf}{\oldincludegraphics[#1]{#2.pdf}}{\IfFileExists{#2.eps}{\oldincludegraphics[#1]{#2.eps}}{\fbox{Missing: \detokenize{#2}}}}}}
\makeatother

\def\codename{\textsc{Praetor}}

\AtBeginDocument{
  \providecommand{\checkmark}{\ding{51}}
  \providecommand{\xmark}{\ding{55}}
}

\begin{document}

\title[Enforcing Benign Trajectories]{Enforcing Benign Trajectories: A Behavioral Firewall for Structured-Workflow AI Agents}

\author*{\fnm{Hung} \sur{Dang} (ORCID: 0009-0008-7009-2509)}
\affil{\orgname{Van Lang University}, \city{Ho Chi Minh City}, \country{Vietnam} \vspace{-10pt}}
\email{hung.dk@vlu.edu.vn}


\abstract{Structured-workflow agents driven by large language models execute tool calls against sensitive external environments. Existing pre-execution firewalls secure these interactions by evaluating each call in isolation. While practical for known signatures, stateless inspection remains vulnerable to context-sequential injection attacks, wherein an adversary navigates through individually benign, syntactically valid tool calls to achieve a malicious objective.  To address this threat, we propose \codename, a telemetry-driven behavioral anomaly detection firewall. Drawing on sequence-based intrusion detection, \codename\ compiles verified benign tool-call telemetry into a parameterized deterministic finite automaton (pDFA). The model defines permitted tool sequences, sequential contexts, and parameter bounds. At runtime, a lightweight gateway enforces these boundaries via an $O(1)$ state-transition structural lookup, shifting computationally expensive analysis entirely offline.  Evaluated on the Agent Security Bench (ASB), \codename\ achieves a 5.6\% macro-averaged attack success rate (ASR) across five scenarios. Within three structured workflows, ASR drops to 2.2\%, outperforming Aegis, a state-of-the-art stateless scanner, at 12.8\%. \codename\ achieves 0\% ASR on multi-step and context-sequential attacks in structured settings. Furthermore, against 1,000 algorithmically spliced exfiltration payloads, only 1.4\% matched valid structural paths, all of which failed end-to-end string parameter guards (0 successes out of 14 surviving paths, 95\% CI [0\%, 23.2\%]). \codename\ introduces just 2.2~ms of per-call latency (a 3.7$\times$ speedup over \textsc{Aegis}) while maintaining a 2.0\% benign task failure rate (BTFR) on benign workloads. Modeling the behavioral trajectory effectively collapses the available attack surface, but unmaintained continuous parameter bounds remain vulnerable to synonym-substitution attacks (18\% evasion rate). Thus, exact-match whitelisting of sensitive parameters ultimately bears the final defensive load against execution.  }

\maketitle

\vspace{-6mm}
\section{Introduction}
\label{sec:intro}

Large language model-driven agents have transitioned to production deployments,
including sensitive domains such as clinical reasoning workflows~\cite{AgenticWorkflow2026},
operating as autonomous principals that invoke APIs and interact with external
services via protocols like the Model Context Protocol (MCP)~\cite{MCP}.
This autonomy expands the attack surface: a single crafted prompt can instruct
an agent to exfiltrate sensitive records or escalate privileges~\cite{AGENTSEC,THREATS}.

To address this, pre-execution firewalls intercept tool calls before they reach the
execution layer~\cite{AEGIS,MSFW,LLMFW}. These systems perform content-first risk
scanning and schema validation. While existing firewalls intercept known signatures~\cite{AEGIS},
they suffer from a fundamental limitation: they evaluate each call as an
isolated event without sequential context. Such sequential blindness creates a systemic vulnerability to \emph{context-sequential injection attacks}, wherein an adversary navigates through a series of individually benign, syntactically valid tool calls to achieve a malicious objective (e.g., exfiltrating data via a seemingly normal email workflow after reading a database). As demonstrated in our diagnostic gap analysis (Section~\ref{sec:eval:rq1}), such attacks bypass state-of-the-art stateless firewalls like \textsc{Aegis} with up to a 75\% success rate, for no single action violates the stateless policy. A stateless firewall examines calls in isolation; hence, it cannot distinguish a legitimate action from an out-of-profile exploit that employs valid parameters.

To mitigate this threat, we propose \codename, a telemetry-driven behavioral anomaly detection firewall for autonomous AI agents. Our approach draws on sequence-based intrusion detection~\cite{FORREST96}.
During the profiling phase, \codename\ ingests a corpus of verified benign tool-call
traces and compiles them into a parameterized deterministic finite automaton (pDFA) that encodes the agent's
expected behavioral envelope. Each state captures a sequential context derived from
the $w$-gram prefix of preceding tool names, and each transition encodes permitted
parameter value bounds.
At runtime, the gateway incurs negligible overhead by performing a constant-time
state-transition lookup against this pre-compiled automaton. Any tool call lacking
a valid outgoing transition is deterministically halted and recorded in a cryptographic
audit log. Consequently, an exploit must produce a sequence indistinguishable from the agent's
established trajectory at every step.

Three properties distinguish \codename\ from existing defenses. First, the gateway is stateful: every decision depends on the session's prior call history rather than on the current call alone. Second, the behavioral profile is per-deployment, compiled from the telemetry of the specific agent it protects and not shared across deployments. Third, the runtime path performs no inference, for all semantic analysis is completed during the offline profiling stage. These three properties permit deployment as a sidecar without modifying the agent's reasoning loop.

Our contributions are as follows:

\begin{itemize}
\item We formalize the problem of context-aware tool-call interception, wherein we define a threat model and establish security goals for bounded-time enforcement (Section~\ref{sec:background}).
\item We design \codename, a decoupled architecture that compiles benign telemetry into a pDFA and enforces strict behavioral boundaries at runtime in constant time (Section~\ref{sec:design}).
\item We evaluate \codename\ on the Agent Security Bench (ASB). The system achieves a \textbf{5.6\%} macro-averaged ASR across five scenarios. Restricted to three structured scenarios, ASR falls to \textbf{2.2\%} (baseline \textsc{Aegis}: \textbf{12.8\%}). The gateway maintains \textbf{2.2}\,ms per-call overhead and a \textbf{2.0\%} benign task failure rate (BTFR) (Section~\ref{sec:eval}).
\end{itemize}
 
\section{Background and Threat Model}
\label{sec:background}

\subsection{LLM Agent Architecture and Tool Calling}
\label{sec:background:arch}

A tool-augmented language model agent operates as a reasoning loop in which
the model receives a system prompt defining its role and constraints, processes
a stream of user inputs and environmental observations, and emits structured
action outputs that are routed to external services~\cite{SURVEY-LLM}.
Each action takes the form of a tool call: a structured invocation consisting
of a tool name and a parameter object conforming to a declared schema.
The agent framework dispatches the call to the appropriate service, receives
the result as an observation, and feeds it back to the model for subsequent
reasoning.
This loop continues until the agent either produces a terminal response or
exhausts its context budget.

The Model Context Protocol (MCP) has emerged as a widely adopted standard for
this dispatch layer, defining a transport-agnostic protocol through which
agents discover available tools and submit invocations~\cite{MCP}.
Agents built on frameworks such as LangChain, CrewAI, and the Anthropic
SDK interact with MCP-compatible servers that expose databases, file systems,
web browsers, and domain-specific APIs as addressable tools.
Since the agent's output is the sole input to these dispatches, and the model
is an untrusted component that may be manipulated via prompt injection, the
tool-call boundary is the natural enforcement point for a pre-execution
security layer.

\begin{table}[t]
\centering
\small
\caption{Summary of notation used in this paper.}
\label{tab:notation}
\begin{tabular}{@{}c p{0.75\columnwidth}@{}}
\toprule
\textbf{Symbol} & \textbf{Description} \\
\midrule
$\mathcal{C}$ & Benign trace corpus \\
$\tau$ & Execution trace \\
$\mathcal{T}$ & Tool vocabulary \\
$\mathbf{p}$ & Parameter vector \\
$\mathbf{P}$ & Parameter space \\
$w$ & Context window width \\
$\theta$ & Pruning threshold \\
$\varepsilon_{\text{num}}, \varepsilon_{\text{str}}$ & Numeric slack factor and string radius expansion \\
$M$ & Parameterized DFA (pDFA) \\
$S, \delta, s_0$ & States, transition function, and initial state \\
$\bot$ & Distinguished idle symbol \\
$\bar{\mathbf{e}}, r$ & Centroid and radius for string parameter guards \\
$Q, p_{\mathrm{pass}}, \delta_{\mathrm{guard}}$ & Oracle query budget, plausible pass probability, and $\Pr[\text{malicious parameter lies outside } r\text{-ball}]$ \\
\bottomrule
\end{tabular}
\end{table} 
\subsection{System Model}
\label{sec:background:model}

We define our system as comprising three entities.

\begin{itemize}

\item \textit{Agent} $\mathcal{A}$: a language model-driven process that
  generates sequences of tool calls based on a fixed system prompt $\Pi$
  and a stream of environmental observations.

\item \textit{Environment} $\mathcal{E}$: the collection of external services,
  databases, and APIs that $\mathcal{A}$ is authorised to invoke.

\item \textit{Gateway} $\mathcal{G}$: the enforcement component interposed
  between $\mathcal{A}$ and $\mathcal{E}$, responsible for allowing or halting
  each tool call before it reaches $\mathcal{E}$.

\end{itemize}

During operation, the agent's behavior manifests as an execution trace. \\

\begin{definition}[Execution Trace]
An \emph{execution trace} is a finite sequence
$\tau = \langle (t_1, \mathbf{p}_1),\, (t_2, \mathbf{p}_2),\, \ldots,\,
(t_k, \mathbf{p}_k) \rangle$, where $t_i \in \mathcal{T}$ is a tool name
drawn from the agent's declared tool vocabulary $\mathcal{T}$, and
$\mathbf{p}_i$ is the parameter vector supplied to tool $t_i$ at step $i$. \\
\end{definition} 

\begin{definition}[Benign Trace Corpus]
The \emph{benign trace corpus} $\mathcal{C} = \{\tau_1, \tau_2, \ldots,
\tau_N\}$ is a set of execution traces collected during verified normal
operation of $\mathcal{A}$, in which no adversarial inputs were present. \\
\end{definition} 

The gateway $\mathcal{G}$ receives each incoming call $(t_i, \mathbf{p}_i)$
in sequence and decides, before forwarding to $\mathcal{E}$, whether the call
is consistent with the agent's established behavioral envelope.
A key design constraint is that $\mathcal{G}$ must reach this decision in
bounded time and without querying any external service, for external queries
would introduce unbounded latency into the agent's execution path.

\subsection{Threat Model and Adversarial Capabilities}
\label{sec:background:threat}

We adopt the indirect prompt injection threat model, wherein the
adversary $\mathcal{X}$ does not have direct access to the agent's system
prompt or model weights but can write content to data sources that the agent
consumes as observations~\cite{THREATS,AGENTSEC}.
Representative injection channels include: crafted entries in a customer
support database that the agent reads, malicious content in a web page the
agent browses, and manipulated tool results returned by a compromised upstream
service. \\

\noindent\textit{Adversary goal.}
$\mathcal{X}$ seeks to cause $\mathcal{A}$ to emit a trace
$\tau_{\mathrm{mal}}$ such that the resulting state change in $\mathcal{E}$
achieves a malicious objective, including data exfiltration, privilege
escalation, or persistent state corruption, while avoiding detection by
whatever security layer is in place. \\

\noindent\textit{Adversary capabilities.}
We define $\mathcal{X}$'s capabilities as follows:

\begin{itemize}

\item {Black-Box input access}: $\mathcal{X}$ can write to any data source
  that $\mathcal{A}$ reads as part of normal operation.

\item {Full knowledge of generic defenses}: $\mathcal{X}$ knows the
  detection patterns, schema validators, and heuristics employed by
  standard, publicly documented firewalls such as \textsc{Aegis}~\cite{AEGIS}.

\item {Observable benign behavior}: $\mathcal{X}$ may, over time,
  observe the agent's benign outputs and thereby form a partial reconstruction
  of its behavioral envelope. \\

\end{itemize}

\noindent\textbf{Adversary limitations.}
We consider two variations of adversarial knowledge. 
In the Black-Box threat model, $\mathcal{X}$ does not possess the specific pDFA compiled by \codename, nor direct access to the historical telemetry corpus $\mathcal{C}$ used to construct it. 
In the Gray-Box threat model, $\mathcal{X}$ has acquired partial access to historical benign traces (e.g., via prolonged observation) and attempts to reconstruct the pDFA constraints. 

We explicitly assume the offline profiling environment is secure and the initial training corpus $\mathcal{C}$ is verified benign (e.g., generated via curated benchmark datasets or collected within a sandboxed, clean-room staging environment). 
Defending against telemetry poisoning during the offline phase (where an adversary actively contaminates the training corpus before the pDFA is compiled) is a critical deployment constraint in behavioral anomaly detection, often referred to as the ``cold start'' problem. This vulnerability surface falls outside the scope of \codename's runtime enforcement guarantees. The security of \codename\ relies on a dual-constraint architecture: structural sequence matching severely collapses the available attack surface, but tight parameter bounds bear the primary load of preventing exploitation within that surviving envelope. We discuss the boundaries and mitigations of this profiling threat surface in Section~\ref{sec:discussion:drift}.

We consider denial-of-service attacks against $\mathcal{G}$ itself, attacks
that bypass the SDK interception layer entirely (e.g., direct API calls), and
compromises of the asynchronous profiling environment to be outside the scope of
this work.

\subsection{Security Goals}
\label{sec:background:goals}

\codename\ must satisfy five properties.

\begin{enumerate}

\item {Soundness.} Any tool call that is inconsistent with the
  $w$-gram behavioral envelope of $\mathcal{C}$, or whose parameter values
  fall outside the $\varepsilon_{\text{num}}$- and $\varepsilon_{\text{str}}$-widened bounds observed in $\mathcal{C}$,
  must be halted before execution.
  Formally, the gateway must strictly enforce a language of permitted traces
  $\mathcal{L}_{\mathrm{permit}}$ that is a deliberate over-approximation of
  $\mathcal{L}(\mathcal{C})$ (the set of tool-call sequences observed in $\mathcal{C}$): traces that are structurally consistent with
  $\mathcal{C}$ but not literally present in it are admitted, reducing false
  positives at the cost of a finite within-profile exploit surface.
  Any trace outside $\mathcal{L}_{\mathrm{permit}}$ is blocked.

\item {Efficiency.} The per-call overhead introduced by $\mathcal{G}$
  must be bounded by a constant with respect to the number of pDFA states,
  the length of the ongoing trace, and the complexity of the agent's tool
  vocabulary.

\item {Tamper-Evidence.} Every blocked call must be recorded in an
  append-only cryptographic audit log such that post-hoc deletion or
  modification of any entry is detectable against any attacker without host-level overwrite capabilities, and globally detectable once published to the external transparency log.

\item {Minimal Integration Burden.} $\mathcal{G}$ must operate as a
  black-box sidecar: it intercepts calls at the tool-dispatch boundary
  and requires no modification to $\mathcal{A}$'s internal reasoning loop,
  system prompt, or model weights.

\item {Evolvability.} The system must support incremental updates to
  the behavioral profile to accommodate legitimate changes in agent tools or
  tasks (concept drift) without requiring a full system re-profiling 
  for every minor update.

\end{enumerate}

\subsection{Motivating Example}
\label{sec:background:example}

Consider a customer-service agent restricted to three tools: \texttt{read\_ticket}, \texttt{write\_summary}, and \texttt{send\_email}. Benign traces follow predictable patterns, e.g., $\langle \texttt{read\_ticket}, \texttt{write\_summary}, \texttt{send\_email} \rangle$. Suppose an adversary injects a prompt instructing the agent to skip the summary phase and jump straight to \texttt{send\_email(to='attacker@external.com')} immediately after reading the ticket. A stateless firewall analyzing these calls in isolation may permit them, as the email format is technically valid and the tool is within the agent's vocabulary. \codename, however, enforces a parameterized deterministic finite automaton (pDFA) where each state encapsulates the sequential context of prior calls, and transitions are guarded by learned parameter bounds (e.g., tight semantic embedding clusters for strings or categorical whitelists). When evaluating the exploit, \codename\ observes that \texttt{send\_email} never appears as a valid outgoing transition directly from the \texttt{read\_ticket} state. The call is halted instantly. Even if the adversary strictly uses authorized tools, they must satisfy both the strict sequential transitions and the tight parameter guards, forcing them to mimic legitimate workflows and neutralizing arbitrary action sequences.
 
\section{Related Work}
\label{sec:related}

\textbf{Stateless Firewalls and the Execution Gap.} Pre-execution firewalls attempt to intercept malicious tool calls before execution. \textsc{Aegis}~\cite{AEGIS} employs a multi-stage pipeline (PII redaction, regex scanning, statistical anomaly detection) but remains fundamentally stateless: it evaluates each call in isolation. Such sequential blindness prevents distinguishing an authorized administrator from an agent tricked into executing a privileged sequence that conforms to static schemas. Other input-layer filters, such as PromptArmor~\cite{PROMPTARMOR} or auxiliary LLM validators~\cite{MSFW,LLMFW}, attempt to sanitize prompts. However, they operate above the tool-call boundary and incur prohibitive inference latency. Moreover, they remain vulnerable to evasion, for adversaries frequently deploy obfuscated injections that bypass prompt-level semantics~\cite{THREATS,LLMAttackSurvey2025}. \codename\ directly addresses this gap by decoupling the security layer into a compiled stateful automaton, enforcing the strict sequential context of prior calls with $O(1)$ constant-time latency. Table~\ref{tab:comparison} summarizes these distinctions. \\

\begin{table}[t]
\centering
\caption{Architectural comparison of \codename\ with representative prior systems.}
\label{tab:comparison}
\setlength{\tabcolsep}{3pt}
\begin{tabular}{@{}lccccc@{}}
  \toprule
  \textbf{System} &
  \textbf{Sequential} &
  \textbf{Agent-} &
  \textbf{Rule/} &
  \textbf{LM/ML} &
  \textbf{Attack} \\
  &
  \textbf{State} &
  \textbf{Specific} &
  \textbf{Pattern} &
  \textbf{Inference} &
  \textbf{Level} \\
  \midrule
  Static Schema                     & \xmark & \xmark & \checkmark & \xmark     & Tool-call \\
  Keyword Filter                    & \xmark & \xmark & \checkmark & \xmark     & Message   \\
  \textsc{Aegis}~\cite{AEGIS}       & \xmark$^a$ & \xmark & \checkmark & \xmark     & Tool-call \\
  MSFW~\cite{MSFW}                   & \xmark & \xmark & \xmark     & \checkmark & Message   \\
  PromptArmor~\cite{PROMPTARMOR}     & \xmark & \xmark & \xmark     & \checkmark & Prompt    \\
  Fides~\cite{FIDES}$^b$       & \xmark & \xmark & \checkmark & \xmark     & Plan/Code \\
  AegisUI~\cite{AEGISUI}             & \xmark & \xmark & \xmark     & \checkmark & Tool-call \\
  \midrule
  \textbf{\codename\ (ours)} & \checkmark & \checkmark & \xmark & \xmark & Tool-call \\
  \bottomrule
  \multicolumn{6}{@{}p{0.95\columnwidth}@{}}{\footnotesize $^a$ \textsc{Aegis} maintains a statistical baseline for event frequencies, but lacks per-session sequential state constraints.}\\
  \multicolumn{6}{@{}p{0.95\columnwidth}@{}}{\footnotesize $^b$ Fides relies on strict taint-tracking labels across variables rather than regex/schema policies.}
\end{tabular}
\end{table} 
\textbf{Stateful Tracking and Intrusion Detection.} The architectural necessity of stateful enforcement has spurred recent concurrent preprints like \textsc{AegisUI}~\cite{AEGISUI}. These systems track execution histories using secondary auditor models or supervised classifiers. While effective, neural inference on the critical path introduces heavy overhead. Alternatively, information-flow control systems like Fides~\cite{FIDES} use dynamic taint tracking to prevent data leakage~\cite{BELLLAPADULA,DENNING76}, but require deep instrumentation of the agent's internal reasoning loop.

\codename's approach is conceptually rooted in classic host-based intrusion detection systems (HIDS). Forrest et al.~\cite{FORREST96} first demonstrated that normal UNIX process behavior can be characterized by short sequences of system calls ($n$-grams) and deviations indicate exploitation~\cite{WARRENDER99}. Recent theoretical work establishing that bounded-context LLM agents are equivalent to probabilistic automata~\cite{AUTOMATA} validates the application of these techniques to modern AI agents. \codename\ adapts this foundational IDS principle to the domain of LLM tool calling, acting as a runtime safety monitor~\cite{LEUCKER09}. However, standard syscall monitoring only inspects control-flow; LLM tool calls carry semantic parameters with high inter-instance variance. \codename\ extends classic $n$-gram tracking by compiling a Symbolic Finite Automaton (SFA)~\cite{SFA} that couples structural transitions with continuous parameter bounding, achieving control-flow and data-flow integrity simultaneously without requiring access to the agent's internals.
 
\section{\codename}
\label{sec:design}

\codename\ separates enforcement into two stages: an asynchronous
\emph{profiling stage} that extracts behavioral norms from historical data,
and a synchronous \emph{runtime enforcement stage} that monitors every
incoming tool call.

During the profiling stage, \codename\ ingests verified benign execution traces $\mathcal{C}$ collected from the agent's normal operations. It compiles this telemetry into a compact, agent-specific behavioral profile. We adopt the formalism of \emph{Symbolic Finite Automata} (SFA)~\cite{SFA} to represent this profile, which we refer to as a \emph{pDFA} (parameterized deterministic finite automaton). The pDFA encodes the agent's expected behavioral envelope: which tool sequences are permitted and within what parameter value bounds. By using a symbolic representation, \codename\ can efficiently represent constraints over complex and infinite parameter domains (e.g., continuous numeric ranges or high-dimensional text embeddings) without state explosion.

During the runtime stage, the Runtime Gateway enforces the pre-compiled pDFA as a transparent sidecar. For each incoming tool call, the gateway performs a constant-time transition lookup. If the call matches a permitted transition from the current state and satisfies the associated parameter guards, it is forwarded to the environment; otherwise, it is halted and logged. The architecture ensures that the computationally expensive task of behavioral analysis is performed offline, while runtime enforcement remains deterministic and lightweight.

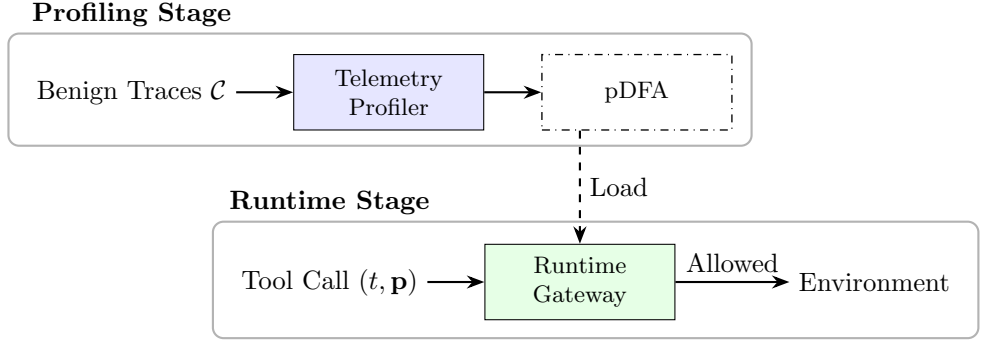
\begin{figure}[t]
\centering
\begin{tikzpicture}[
    node distance=1.5cm,
    box/.style={rectangle, draw, minimum width=2.5cm, minimum height=1cm, text centered, font=\small},
    arrow/.style={-Stealth, thick}
]
\node (telemetry) {Benign Traces $\mathcal{C}$};
    \node (profiler) [box, right=0.75cm of telemetry, fill=blue!10, text width=1.75cm, align=center] {Telemetry Profiler};
    \node (pdfa) [box, right=0.75cm of profiler, dash dot, line width = 0.5pt] {pDFA};
    
\node (gateway) [box, below=1.5cm of pdfa, xshift=-0.75cm, fill=green!10, text width=1.75cm, align=center] {Runtime Gateway};
    \node (call) [left=0.75cm of gateway] {Tool Call $(t, \mathbf{p})$};
    \node (env) [right=1.5cm of gateway] {Environment};
    
    \draw [arrow] (telemetry) -- (profiler);
    \draw [arrow] (profiler) -- (pdfa);
    \draw [arrow] (call) -- (gateway);
    \draw [arrow] (gateway) -- (env) node[midway, above] {Allowed};
    \draw [arrow, dashed] ($(pdfa.south) + (-0.75cm, 0)$)-- (gateway) node[midway, right] {Load};
    
    \draw [draw=black!30, thick, rounded corners] ($(telemetry.north west)+(-0.25,0.5)$ ) rectangle ($(pdfa.south east)+(0.25,-0.2)$);
    \node at ($(telemetry.north)+(0,0.75)$) {\textbf{Profiling Stage}};

    \draw [draw=black!30, thick, rounded corners] ($(call.north west)+(-0.25,0.5)$ ) rectangle ($(env.south east)+(0.25,-0.5)$);
    \node at ($(call.north)+(0,0.75)$) {\textbf{Runtime Stage }};

\end{tikzpicture}
\caption{The two-stage lifecycle of \codename. The telemetry profiler compiles historical telemetry into a pDFA, which the runtime gateway enforces at execution time with $O(1)$ overhead.}
\label{fig:architecture}
\end{figure}

Figure~\ref{fig:architecture} illustrates the two-stage pipeline.

Our design assumes narrow-task, single-role agents, the
primary deployment target of \codename, where a small, stable tool
vocabulary and a well-defined task scope produce low-entropy call sequences.
Agents with large tool vocabularies, open-ended task distributions, or
dynamically injected context may exhibit higher behavioral entropy, which
would widen the pDFA's state space and increase the benign task failure rate
for a fixed pruning threshold $\theta$; we discuss this scope limitation
in Section~\ref{sec:discussion:entropy}.

\subsection{Telemetry Profiler}
\label{sec:design:profiler}

The Telemetry Profiler takes as input a benign trace corpus and three parameters which are a context-window width $w$, a pruning threshold $\theta$ and slack factors $\varepsilon_{\text{num}}, \varepsilon_{\text{str}}$, wherein

\begin{itemize}
\item A corpus of verified benign traces
  $\mathcal{C} = \{\tau_1, \ldots, \tau_N\}$, where each
  $\tau_i = \langle (t_1, \mathbf{p}_1), \ldots, (t_k, \mathbf{p}_k) \rangle$.
\item A context-window width $w$ (default $w = 3$), controlling the length of
  the sequential prefix used to distinguish states with the same tool name but
  different prior histories.\footnote{While $w=3$ is selected as the optimal operating point for our evaluated scenarios, this parameter is deployment-dependent: agents with higher behavioral entropy or more complex branching may require larger windows to achieve sufficient precision.}
\item A pruning threshold $\theta$ (default $\theta = 3$), specifying the
  minimum number of observed transitions required to retain a state.
\item Slack factors $\varepsilon_{\text{num}}$ (default $0.05$) and $\varepsilon_{\text{str}}$ (default $0.05$),
  widening numeric parameter bounds and string embedding radii to accommodate minor benign variation.
\end{itemize}

It transforms the input described above into a pDFA following the four steps described below.  \\

\begin{figure}[t]
\centering
\resizebox{0.95\columnwidth}{!}{\begin{tikzpicture}[
    node distance=0.5cm,
    trace_box/.style={rectangle, draw=black!30, fill=white, minimum width=1.5cm, minimum height=0.6cm},
    window/.style={draw=blue!80,  rounded corners, line width=1pt, opacity=0.8},
    state_box/.style={rectangle, draw=blue!80, fill=blue!10, rounded corners, minimum width=3cm, minimum height=1.2cm,  align=center},
    arrow/.style={-Stealth, thick}
]
\node[trace_box] (t1) {read};
    \node[trace_box, right=of t1] (t2) {write};
    \node[trace_box, right=of t2] (t3) {send};
    \node[trace_box, right=of t3] (t4) {close};
    
    \node[font=\sffamily\bfseries, left=0.25cm of t1.west, text width=1.75cm, align=center] {Execution Trace $\tau$};

\node[window, fit=(t1) (t2), inner sep=4pt, dashed, fill=none] (w1) {};
    \node[blue!80, above=0.1pt of w1] {$w$-gram Context};

\node[state_box, below=1.5cm of t3.south west] (s_map) {
        \textbf{State Abstraction} \\
        $s = (t_j, \mathrm{ctx}_j)$ \\
        $s = (\text{\texttt{send}}, [\text{\texttt{read, write}}])$
    };

    \draw[arrow] (w1.south) -- (s_map.north west) node[midway, left] {History};
    \draw[arrow] (t3.south) -- (s_map.north) node[midway, right] {Current Tool};

\node[state_box, right=1.5cm of s_map] (s_next) {
        $s' = (\text{\texttt{close}}, [\text{\texttt{write, send}}])$
    };
    \draw[arrow, black] (s_map) -- (s_next) node[midway, above] {$\delta_{\mathrm{struct}}$};

\node[purple, below=0.2cm of s_map] {$\mathrm{ctx}_j = (t_{j-w}, \dots, t_{j-1})$};

\end{tikzpicture} }
\caption{The state abstraction mechanism. A sliding $w$-gram window (blue) captures the tool name sequence preceding the current call. This context is concatenated with the current tool name to form a unique pDFA state $(t_j, \mathrm{ctx}_j)$. Sequential transitions are then mapped to the symbolic transition function $\delta_{\mathrm{struct}}$.}

\label{fig:state_abstraction}
\end{figure}
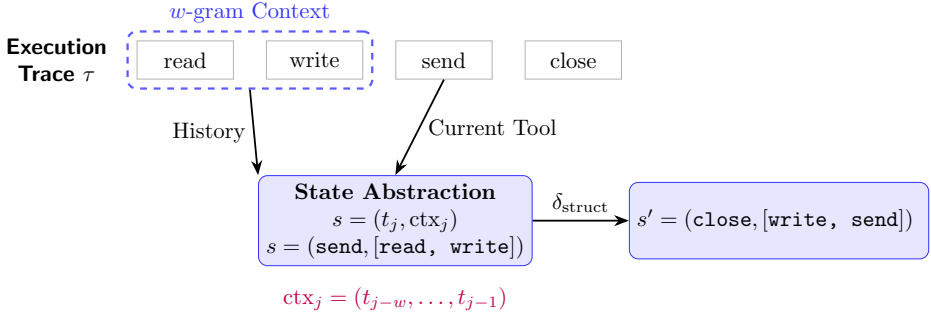

\textit{Step 1: State Abstraction.}
Each unique combination of a tool name and a context prefix defines a distinct
state.
Formally, for each position $j$ in each trace $\tau_i$, the context is
$\mathrm{ctx}_j = (t_{\max(1,\,j-w)}, \ldots, t_{j-1})$, i.e., the
$\min(j{-}1,\,w)$-length prefix of tool names immediately preceding call $j$
(empty for $j = 1$).
The corresponding state is the pair $s_j = (t_j,\, \mathrm{ctx}_j)$,
with $t_j \in \mathcal{T}$ and $\mathrm{ctx}_j \in \mathcal{T}^{\le w}$.
As visualized in Figure~\ref{fig:state_abstraction}, two calls to the same tool in different sequential contexts yield distinct
states, enabling the automaton to distinguish, for example, a
\texttt{read\_ticket} following a session start from one following a
\texttt{write\_summary}. When an agent exhibits iterative error-correction loops or ReAct retries (e.g., repeatedly calling \texttt{search\_web}), these sequences naturally form valid cyclic paths or self-edges within the graph. The pDFA therefore gracefully accommodates expected hallucination or retry patterns without state fragmentation. \\

\textit{Step 2: Transition Extraction.}
For each consecutive pair of states $(s_j, s_{j+1})$ within any trace, the
profiler records the directed edge $s_j \to s_{j+1}$ and associates the
parameter vector $\mathbf{p}_{j+1}$ with that edge.
After processing all traces, each edge accumulates a multiset of observed
parameter vectors. \\

\textit{Step 3: Parameter Schema Synthesis.}
For each edge $e = (s, s')$ in the transition set, the profiler synthesises a
parameter schema from the accumulated observations.
Three parameter types are handled:

\begin{itemize}

\item \emph{Numeric parameters}: let $\{x^{(1)}, \ldots, x^{(m)}\}$ be the
  observed values for parameter dimension $d$.
  The schema records the widened interval
  $[\min_j x^{(j)} - \varepsilon_{\text{num}} \cdot |\min_j x^{(j)}|,\; \max_j x^{(j)} + \varepsilon_{\text{num}} \cdot |\max_j x^{(j)}|]$.
  An incoming value $v$ is accepted if and only if $v$ lies within this interval.

\item \emph{String parameters}: each observed string value $s^{(j)}$ is
  embedded using a sentence transformer (specifically, \texttt{all-MiniLM-L6-v2}
  from the \texttt{sentence-transformers} library) to produce a fixed-dimensional
  vector $\mathbf{e}^{(j)} \in \mathbb{R}^d$.
  The schema records the L2-normalized centroid $\bar{\mathbf{e}} = \frac{1}{m}\sum_j
  \mathbf{e}^{(j)}$ and the radius $r = \max_j D_C(\mathbf{e}^{(j)}, \bar{\mathbf{e}}) + \varepsilon_{\text{str}}$, where $D_C$ is the cosine distance.
  An incoming string is accepted if its cosine distance to $\bar{\mathbf{e}}$ is at most $r$.
  Embeddings are computed once at profiling time and cached; the runtime gateway embeds novel incoming strings using a quantized transformer.
  To counter semantic evasion (Section~\ref{sec:eval}), the profiler applies a \textbf{Sensitive-Parameter Whitelist Override}: parameters matching predefined sensitive patterns (e.g., \texttt{*\_path}) bypass continuous bounds and enforce strict categorical exact-matching against the observed corpus.

\item \emph{Categorical parameters}: the schema records the exact set of
  observed values $\mathcal{V}_d = \{v^{(1)}, \ldots, v^{(m)}\}$.
  An incoming value is accepted only if it belongs to $\mathcal{V}_d$. \\

\end{itemize}

\textit{Step 4: DFA Construction and Pruning.}
(Note: For conceptual clarity, we present schema synthesis before pruning in this exposition; an optimized implementation applies pruning first, as reflected in Algorithm~\ref{alg:profiler}).
The profiler assembles the pDFA $M = (S, \delta, s_0)$ whose transition
function decomposes into two orthogonal components.

\noindent\emph{Structural successor.}
Define $\delta_{\mathrm{struct}}: S \times \mathcal{T} \rightharpoonup S$ as
the partial function that maps a state $s = (t_j, \mathrm{ctx}_j)$ and a
tool name $t_{\mathrm{next}}$ to the unique successor
$s' = \bigl(t_{\mathrm{next}},\; (t_{\max(1,j-w+1)}, \ldots,
t_j)\bigr)$, i.e., the state whose tool component is $t_{\mathrm{next}}$ and
whose context component is the $w$-gram formed by appending $t_j$ to the
tail of $\mathrm{ctx}_j$.
This successor is fully determined by $s$ and $t_{\mathrm{next}}$ alone;
$\delta_{\mathrm{struct}}$ is therefore structurally deterministic, with at
most one outgoing edge per $(s, t_{\mathrm{next}})$ pair.
$\delta_{\mathrm{struct}}(s, t_{\mathrm{next}})$ is undefined if no trace
in $\mathcal{C}$ ever produced the transition $(s \to s')$.

\noindent\emph{Parameter guard.}
For each defined edge $(s \to s')$, the predicate
$\mathrm{guard}(s, t_{\mathrm{next}}, \mathbf{p})$ evaluates to $\mathbf{true}$
if and only if $\mathbf{p}$ satisfies the parameter schema synthesised for
that edge in Step~3.

\noindent\emph{Combined transition.}
The full transition function $\delta: S \times (\mathcal{T} \times \mathbf{P})
\rightharpoonup S$ is:
$$
\delta\!\bigl(s,\,(t,\,\mathbf{p})\bigr) \;=\;
\begin{cases}
\delta_{\mathrm{struct}}(s, t)
& \text{if } \delta_{\mathrm{struct}}(s, t) \text{ is defined}
  \text{ and } \mathrm{guard}(s, t, \mathbf{p}) = \mathbf{true}\\
\text{undefined} & \text{otherwise.}
\end{cases}
$$
For any fixed $(s, t)$, $\delta_{\mathrm{struct}}$ yields at most one
structural successor, and there is exactly one guard predicate per $(s, t)$
pair; the automaton is therefore \emph{deterministic} in the sense required
by SFA theory~\cite{SFA}: no two distinct outgoing transitions from $s$ can
be satisfied by the same concrete input $(t, \mathbf{p})$.

\noindent\emph{Pruning.}
States with fewer than $\theta$ total observed outgoing-transition instances
($\sum_{s'} \mathit{edges}[s, s'] < \theta$) are pruned from $S$, for
rarely-visited states yield unreliable schema estimates.
After each removal, all edges whose source \emph{or destination} is the
pruned state are deleted from the transition relation; states that thereby
lose all their observations may themselves fall below threshold, so the
procedure iterates to a fixpoint, analogously to dead-state elimination
in standard DFA minimisation. Finally, a forward reachability pass from $s_0$ is 
performed to remove any states that became unreachable due to pruned edges.
Because iterative pruning executes to a true fixpoint, no surviving state maintains a structural transition to a pruned state; hence, after pruning, any state that remains structurally reachable from $s_0$ is guaranteed to satisfy the $\theta$ threshold, and a single post-fixpoint reachability pass is sufficient to clear orphaned subgraphs.
The initial state $s_0 = (\bot, \lambda) \in
(\mathcal{T} \cup \{\bot\}) \times \mathcal{T}^{\le w}$ is never pruned.
Making $\delta$ partial eliminates the need for an explicit accept set:
states from which $\delta$ is undefined for every input are terminal, and
the gateway blocks any call emitted while the session pointer is at such a state.
The resulting automaton is serialised to a compact binary representation
(MessagePack) for loading by the runtime gateway.

The core logic of the telemetry profiler is summarised in Algorithm~\ref{alg:profiler}. 

\begin{algorithm}[htbp]
\caption{Telemetry Profiler (Core Logic)}
\label{alg:profiler}
\begin{algorithmic}[1]
\Require Corpus $\mathcal{C}$, context width $w$, threshold $\theta$, slack factors $\varepsilon_{\text{num}}, \varepsilon_{\text{str}}$
\Ensure pDFA $M = (S, \delta, s_0)$
\State $s_0 \leftarrow (\bot, \lambda); S \leftarrow \{s_0\}$
\For{each trace $\tau \in \mathcal{C}$}
  \State $s_{\mathrm{prev}} \leftarrow s_0$
  \For{each call $(t_j, \mathbf{p}_j)$ in $\tau$}
    \State $\mathrm{ctx}_j \leftarrow w$-gram prefix preceding $t_j$
    \State $s_j \leftarrow (t_j, \mathrm{ctx}_j); S \leftarrow S \cup \{s_j\}$
    \State Record edge $(s_{\mathrm{prev}} \to s_j)$ and parameters $\mathbf{p}_j$
    \State $s_{\mathrm{prev}} \leftarrow s_j$
  \EndFor
\EndFor
\State \textbf{Prune \& Reachability:} Iteratively remove states where outgoing edge counts $< \theta$, deleting connected edges. Finally, remove states unreachable from $s_0$.
\State \textbf{Synthesize Schemas:} For each edge $(s \to s')$ in the pruned graph, compute widened numeric bounds ($\pm \varepsilon_{\text{num}}$) and categorical sets. For string parameters, compute SBERT centroids $\bar{\mathbf{e}}$ and radii $r$. Map to $\mathrm{guard}(s, t, \mathbf{p})$.
\State \textbf{Construct $\delta$:} Map $(s, t, \mathbf{p}) \mapsto s'$ iff $\delta_{\mathrm{struct}}(s, t) = s'$ and $\mathrm{guard} = \mathbf{true}$.
\State \Return $M = (S,\, \delta,\, s_0)$
\end{algorithmic}
\end{algorithm}

\textit{Deployment Constraint: Embedding Drift.}
Because string parameter schemas rely on embeddings computed at profiling time, 
if the sentence transformer model (e.g., \texttt{all-MiniLM-L6-v2}) is updated 
or changed between profiling cycles, the embedding space will shift and all cached 
embeddings will become invalid. Re-profiling the corpus is required when the embedding model is updated. \\

\begin{remark}[Determinism]
\label{rem:determinism}
The pDFA $M$ produced by Algorithm~\ref{alg:profiler} is a \emph{deterministic}
SFA: for any state $s \in S$ and any concrete input $(t, \mathbf{p}) \in
\mathcal{T} \times \mathbf{P}$, there is at most one defined value of
$\delta(s, (t, \mathbf{p}))$.
Since each state $s' = (t_{\mathrm{next}}, \mathrm{ctx}')$ is uniquely
determined by the pair $(s, t_{\mathrm{next}})$ via the deterministic context
update, $\delta_{\mathrm{struct}}$ contains at most one edge per
$(s, t_{\mathrm{next}})$ pair.
Because tool names in $\mathcal{T}$ are discrete, two distinct outgoing edges
from $s$ necessarily carry \emph{different} tool-name labels; therefore, guard-predicate overlap across different edges is structurally impossible because they are keyed by distinct $(s, t_{\mathrm{next}})$ pairs.
Within the same $(s, t)$ pair there is exactly one compiled schema and hence
one guard predicate; no input can satisfy two guards simultaneously.
The automaton therefore meets the determinism condition for SFAs stated
in~\cite{SFA}. \\
\end{remark}

\textit{Complexity.}
Let $N$ be the number of traces, $L$ the maximum trace length, and $|\mathcal{T}|$
the tool vocabulary size.
The state abstraction and transition extraction steps run in $O(N \cdot L)$.
Parameter schema synthesis for string parameters requires one embedding
computation per unique string value; embeddings are cached so that identical
strings are embedded only once. In a production implementation, applying pruning first on structural edge counts bounds the waste of synthesizing schemas for eventually discarded edges.
The theoretical maximum state count before pruning is loosely bounded by $|\mathcal{T}|^{w+1}$ (it is a loose upper bound because for any structurally reachable adjacent states $s \to s'$, the last element of $s'$'s context must equal $s$'s tool component). For $|\mathcal{T}| = 15$ and $w = 3$, this bounds the space at 50,625 states. Empirically, the observed pruned state count strictly tightens to 450 (Section~\ref{sec:impl}), producing a flat, memory-resident structure independent of corpus size.
The telemetry profiler runs once per agent deployment or update cycle, not
on the critical path.

\subsection{Runtime State-Machine Gateway}
\label{sec:design:gateway}

At startup, the gateway loads the serialised pDFA into a hash-map-backed
adjacency structure.
It allocates a per-session state pointer, initialised to $s_0$, for each
active agent session. Upon the emission of a tool call $(t, \mathbf{p})$ by the agent, the gateway proceeds
as follows:

\begin{enumerate}

\item \emph{Edge lookup} (O(1)): lookup of the single edge $(s_{\mathrm{cur}}, t)$ from the
  current state $s_{\mathrm{cur}}$ by keying into the adjacency map.
  The total per-call overhead is therefore $O(1)$ in $|\mathcal{T}|$, the tool vocabulary
  size, which is a small constant for practical agents.

\item \emph{Schema check}: for each candidate edge $(s_{\mathrm{cur}}, s')$
  labelled with tool $t$, check whether $\mathbf{p}$ satisfies the
  edge's parameter schema.
  Numeric checks require one comparison per dimension; categorical checks
  require one hash-set lookup; string checks require a forward pass through a quantized embedding model
  and one cosine-similarity comparison against the pre-cached centroid.
  The embedding dimension $d$ is a fixed constant (384 for
  \texttt{all-MiniLM-L6-v2}), so string checks run in $O(d) = O(1)$.

\item \emph{Decision}: if a matching edge exists, the gateway forwards the
  call to the environment and advances the state pointer to $s'$.
  If no matching edge exists, the gateway blocks the call, returns a
  sanitised error message, and appends the event to the cryptographic audit log. The state pointer remains strictly at $s_{\mathrm{cur}}$ upon rejection, enabling benign retries while exposing a static state to adversarial probing. \\

\end{enumerate}

The structural lookup is therefore $O(1)$ in $|\mathcal{T}|$. While overall latency remains bounded by the fixed inference cost of the quantized embedding model ($O(d)$), the architectural path complexity scales as $O(1)$ with respect to the agent's behavioral footprint ($|S|$, $k$, $N$, $L$).
Unlike the pipeline in~\cite{AEGIS}, the gateway performs no string scanning,
no pattern matching, and no policy evaluation at runtime; all such
analysis is precomputed.

\textit{Cryptographic audit log.}
Each blocked event is recorded as a tuple
$\ell = (\mathit{session\_id},\; s_{\mathrm{cur}},\; t,\; \mathbf{p},\;
\mathit{timestamp})$.
The log is maintained as an append-only SHA-256 hash chain: the $n$-th
entry stores both the event data and the hash of the preceding entry,
so that any post-hoc deletion or modification of an entry invalidates all
subsequent chain hashes.
The chain root is published periodically to an append-only transparency
log, enabling external forensic verification without access to the raw
event data.

\subsection{Incremental Profile Updates}
\label{sec:design:updates}

A deployed agent may evolve: new tools may be added, the system prompt may be
revised, or the agent may be retrained on an updated corpus.
If the pDFA is not updated to reflect these changes, the gateway will generate
false positives for legitimate new behaviors.

To address this, we introduce a human-in-the-loop incremental update protocol.
When the gateway blocks a call that is subsequently reviewed and approved by a
human operator, either because it represents a legitimate new workflow or a
new tool, the approved event is appended to a \emph{pending update queue}.
During the next scheduled asynchronous recompilation window (e.g., a nightly pass),
the profiler ingests the pending queue as supplementary trace data and
re-expands the DFA to accommodate the newly approved transitions, without
reprocessing the full historical corpus.
Specifically, only the states and edges reachable from the newly approved
sequences are reconstructed; the remainder of the DFA is preserved.

The protocol preserves two key properties.
First, new behaviors enter the behavioral envelope only after explicit human
approval, preventing the gateway from silently expanding its permissions in
response to adversarial inputs.
Second, the incremental recompilation step runs asynchronously and does not affect
gateway availability.
We evaluate the practical impact of this protocol on false positive rates in
the concept-drift ablation study described in Section~\ref{sec:eval:rq4}.
 
\section{Security Analysis}
\label{sec:security}

The security guarantees of \codename\ reduce to three properties: execution path integrity, computational efficiency, and tamper-evident auditing.

\subsection{Execution Path Integrity}
\label{sec:security:integrity}

Unlike stateless firewalls that evaluate tool calls against static schemas or keyword lists, \codename\ enforces \emph{execution path integrity}. We define the safety of an agent not merely by the validity of its individual actions, but by its conformance to a known-benign trajectory.

Let $\mathcal{T}$ be the agent's tool vocabulary. A stateless firewall monitors the entire transition space $\mathcal{T}$ at every step. In contrast, \codename\ restricts the agent's operational space to the out-degree of its current state $s$ in the pDFA $M$. Formally, at any state $s$, the set of permitted tools is:
\begin{equation}
\mathcal{T}_{\mathrm{permitted}}(s) = \{ t \in \mathcal{T} : \text{the edge } (s \xrightarrow{t} s') \text{ exists in } M \}
\end{equation}
By design, $|\mathcal{T}_{\mathrm{permitted}}(s)| \ll |\mathcal{T}|$ for task-specific agents. For the Customer Service scenario evaluated in Section~\ref{sec:eval}, the average out-degree is $1.8$, while the total vocabulary $|\mathcal{T}| = 15$. Crucially, at states preceding high-value terminal tools, the maximum observed out-degree is strictly bounded to 4. This represents an {88\% average reduction in the immediate attack surface} per step. To execute a malicious action $t_{adv}$, an adversary must not only bypass the parameter guards but must also navigate the agent into a state $s$ where $t_{adv} \in \mathcal{T}_{\mathrm{permitted}}(s)$. Such topological constraints force attackers to mimic benign sequential patterns perfectly. The system thereby neutralizes multi-step exfiltration attacks that rely on ``jump-starting'' sensitive tools from an idle or unrelated context.

\subsection{Adversary Model and Path-Guessing}
\label{sec:security:adversary}

We assume an adversary $\mathcal{X}$ who can induce the agent to emit arbitrary tool calls via prompt injection or tool poisoning. The adversary's objective is to execute a target sequence $\tau_{adv}$ that results in a security breach (e.g., data exfiltration). We distinguish two levels of adversarial knowledge:

\begin{enumerate}
\item {Black-Box Adversary}: Knows the tool vocabulary $\mathcal{T}$ but lacks access to the pDFA $M$, the training corpus $\mathcal{C}$, or the embedding model.
\item {Gray-Box Adversary}: Has access to a subset of historical benign traces $\mathcal{C}' \subset \mathcal{C}$ and attempts to reconstruct the pDFA.
\end{enumerate}

Under the Black-Box model, a competent attacker does not sample paths uniformly; they enumerate sequences reaching high-value terminal tools. Thus, the true attacker advantage is tightly bounded by the fraction of permitted paths terminating in a target tool (e.g., \texttt{read\_db}), relative to all paths reaching that tool. In the Customer Service pDFA, only 3 of the 142 valid paths terminate in \texttt{read\_db} (a measured property of the trained graph), meaning an adversary attempting to jump-start an exfiltration exploit faces a highly constrained maze. On a macro level, this 142-path structural envelope represents extreme global sparsity, covering just 0.018\% of the uniform combinatorial action space for sequences of length $k=5$ ($\rho \approx 1.8 \times 10^{-4}$). To quantify this residual risk empirically, we conducted an algorithmic path enumeration experiment distinct from the full ASB evaluation in Section~\ref{sec:eval}: we generated 1,000 multi-step exfiltration payload traces by algorithmically splicing valid ASB exfiltration templates onto benign prefixes, producing 1,000 syntactically distinct terminal actions reaching sensitive tools, and mapped them against the pDFA structure. Only 14 payloads (\textbf{1.4\%}) contained sequences that coincidentally matched the benign structural paths. However, when executing these 14 surviving paths in an end-to-end simulation with the actual agent, 0 achieved a meaningful exfiltration objective (95\% CI: [0\%, 23.2\%]). The malicious parameter payloads (e.g., specific attacker-controlled email addresses or unauthorized SQL queries) triggered the continuous string parameter guards (the $\varepsilon_{\text{str}}$-ball), resulting in blocks. This demonstrates that structural graph traversal alone is insufficient for exploitation.

For the Gray-Box adversary, the security of \codename\ relies on the
     \emph{unpredictability of continuous parameter guards}. To formalize this, we
     model the attack as a bounded evasion game. Let the firewall be a deterministic
     execution oracle $\mathcal{O}_M$ that evaluates a sequence $\tau$ against the
     stateful pDFA $M$ and returns $1$ (Allow) or $0$ (Block). The adversary
     $\mathcal{A}$, constrained by a budget of $Q$ queries, attempts to construct an
     adversarial trace $\tau^* = \tau_{\mathrm{pre}} \circ (t^*,
     \mathbf{p}_{\mathrm{mal}})$ such that $\mathcal{O}_M(\tau^*) = 1$, where
     $\tau_{\mathrm{pre}}$ is a valid structural prefix and
     $\mathbf{p}_{\mathrm{mal}}$ achieves a targeted malicious objective.
   
    Within this game, we ground two empirical quantities representing the
     adversary's capability. Let $p_{\mathrm{pass}}$ be the probability that a
     benign heuristic parameter (e.g., an empty string or schema default), generated
     by $\mathcal{A}$ to probe structural transitions, satisfies the guard
     conditions. Conversely, let $\delta_{\mathrm{guard}} = \Pr_{\mathbf{p} \sim
     \mathcal{D}_{\mathrm{mal}}}[\mathcal{O}_M(\tau_{\mathrm{pre}} \circ (t^*,
     \mathbf{p})) = 0]$ represent the firewall's expected rejection rate over the
     adversary's distribution of malicious payloads, $\mathcal{D}_{\mathrm{mal}}$.
   
    Under this formulation, the adversary's expected query complexity decomposes
     logically. Because a Block on a heuristic strongly narrows the search space,
     $\mathcal{A}$ leverages an expected $Q \cdot p_{\mathrm{pass}}$ successful
     Allow signals to map the valid structural transitions
     $\mathcal{T}_{\mathrm{permitted}}(s)$ efficiently. Thus, structural opacity
     alone is not a durable defense against a query-rich oracle. However, the
     expected discovery of an \emph{exploitable} parameter payload scales strictly
     with $Q \cdot (1 - \delta_{\mathrm{guard}})$. Here, $\delta_{\mathrm{guard}}$
     is an empirical property of the embedding geometry: because text mutations
     operate over a discrete manifold within a continuous space, gradient-free
     boundary approximation (e.g., HopSkipJump) requires an exhaustive query budget
     to locate an adversarial example within the permitted $r$-ball. Under practical
     rate limits ($Q \le 500$), convergence is computationally prohibitive.
     Consequently, $\delta_{\mathrm{guard}}$ remains near $1$ (empirically $\approx 0.98$), which dominates the adversary's expected discovery rate.

\subsection{Computational Efficiency and $O(1)$ Structural Enforcement}
\label{sec:security:efficiency}

A critical design goal of \codename\ is to maintain negligible runtime overhead. The structural path validation is $O(1)$ with respect to the number of pDFA states, the current trace length, and the original corpus size. This efficiency is achieved because the gateway maintains an explicit state pointer, reducing enforcement to a simple dictionary lookup. The number of outgoing transitions from any state is bounded by the agent's tool vocabulary size, and categorical/numeric parameter schema checks are performed in constant time. While string parameter evaluation requires a fixed-budget forward pass through an embedding model (which dominates the absolute latency, as evaluated in Section~\ref{sec:eval:rq2}), the architectural complexity remains entirely decoupled from the size of the behavioral profile. This structural advantage ensures that \codename\ does not suffer from the performance bottlenecks typical of runtime language-model inference or exhaustive signature matching, which typically scale with the number of rules or patterns.

\subsection{Residual Attack Surface: Within-Profile Exploits}
\label{sec:security:knowledge}

The primary residual threat to \codename\ is the \emph{within-profile exploit}, where an adversary achieves a malicious objective using only legitimate, observed transitions. This is analogous to return-oriented programming (ROP) in traditional software security.

Nonetheless, \codename\ substantially constrains such exploits. The context window prevents stitching unrelated benign fragments, for the adversary must maintain a consistent $w$-gram history throughout the attack. The parameter schema further prevents tool repurposing (e.g., a file-read tool observed only on \texttt{/tmp/logs} cannot be invoked on \texttt{/etc/passwd}), provided the parameter is protected by the Sensitive-Parameter Whitelist Override defined in Section~\ref{sec:design}. These constraints do not eliminate all semantic-gap attacks; they confine the adversary to the agent's observed envelope, within which the capabilities available for exploitation are a small fraction of those an unconstrained agent would expose.
 
\section{Implementation}
\label{sec:impl}

We implement \codename\ as two cooperating processes: a Telemetry Profiler written
in Python and a Runtime Gateway written in Rust. \\

\textbf{Telemetry Profiler.}
The profiler is implemented as a standalone Python module of approximately
{250} lines of code.
It accepts trace logs in JSON Lines format, in which each line encodes one tool
call as a (tool name, parameter object, session ID, timestamp) tuple.
String parameter embeddings are computed using the
\texttt{all-MiniLM-L6-v2} model from the \texttt{sentence-transformers}
library~\cite{SBERT}, which produces 384-dimensional vectors with strong
semantic alignment properties at low inference cost.
Embeddings are cached in a persistent on-disk store keyed by string content,
so that repeated occurrences of the same argument value across traces incur
only a single embedding computation.
The resulting pDFA is serialised to MessagePack, a compact binary format that
avoids the overhead of JSON parsing at gateway startup.
For a representative customer-service agent with {15} tools and
{500} benign traces, the profiler produces a pDFA of
{450} states and {620} edges, with a serialised
binary size of {85}\,KB (as only $\sim$10\% of edges carry dense 384-dimensional string centroids, while the rest use lightweight categorical or numeric schemas) and a total compilation time of 
{12}\,s. 

\textbf{Runtime Gateway.}
The gateway is implemented in Rust as a long-running daemon of approximately
{420} lines of code.
At startup, it loads the MessagePack-serialised pDFA into a
\texttt{HashMap}-backed adjacency structure, allocating one hash map per state
keyed by (tool name, schema-type) pairs.
The gateway exposes a local Unix domain socket API whose request/response
envelope is compatible with both the OpenAI function-calling format and the
Anthropic tool-use API, enabling direct integration without modification to
existing agent code.
Per-session state is maintained in a globally shared, mutex-protected hash map keyed by session ID, safely coordinating concurrent sessions across the asynchronous runtime.
The in-memory footprint of the gateway process for a {15}-tool
customer-service agent is {41}\,MB (this footprint represents the total Resident Set Size of the daemon, including the Rust Tokio asynchronous runtime, HTTP framework buffers, pre-allocated thread pools, and approximately 23\,MB for the \texttt{INT8}-quantized ONNX model; the pDFA itself consumes $< 1$\,MB (accounting for the 10\% of edges carrying 1.5\,KB centroids and the adjacency map overhead)).
Gateway startup latency from process launch to first request served is
{45}\,ms, dominated by loading the transformer weights into the \texttt{ort} ONNX session. 

\textbf{Audit log.}
The SHA-256 hash chain is maintained in an append-only write-ahead log on the
local file system.
Each blocked event is synchronously flushed before the rejection response is
returned to the agent, ensuring that no blocked call can be omitted from the
log due to a crash between the blocking decision and the write.
While the underlying design supports exporting the chain root on a configurable interval to a Sigstore-compatible transparency log, this external integration is omitted from the open-source baseline artifact for simplicity. The interval defines the security parameter for the window of vulnerability against a compromised host attempting to truncate the log; once exported, the chain enables
independent third-party verification of audit completeness. 

\textbf{API compatibility.}
The gateway acts as a transparent middleware layer.
Agent frameworks that support the OpenAI \texttt{tool\_use} response format require only a
one-line configuration change to route tool-call requests through the gateway's
Unix socket before dispatch.
The gateway returns the same response envelope regardless of whether the call
is allowed or blocked: an allowed call is forwarded to the environment and the
result returned verbatim; a blocked call returns an error payload in the same
format, allowing the agent to handle the rejection gracefully within its
existing error-handling logic.
 
\section{Evaluation}
\label{sec:eval}

\subsection{Experimental Setup}
\label{sec:eval:setup}

\textbf{Hardware.}
All experiments are conducted on AWS \texttt{c6i.4xlarge} instances
(16 vCPU, Intel Xeon Ice Lake, 32\,GB RAM), representing a standard
compute-optimised deployment environment.
The \codename\ gateway and \textsc{Aegis} itself (deployed as an out-of-process sidecar) run on
separate instances of the same class to ensure fair latency comparisons. \\

\textbf{Benchmarks.}
Our evaluation draws on two sources: ASB for attack scenarios and ToolBench for benign trace generation.
For evaluation, we use Agent Security Bench (ASB)~\cite{ASB},
which provides ten agent scenarios with over 400 tools and a standardised
attack suite spanning five families: Direct Injection (DI), Indirect Injection (II), Multi-Step Exfiltration (MSE), Tool Poisoning (TP), Memory Manipulation (MM).

While the full ASB benchmark aggregates over 400 tools, production agents operate within tightly scoped personas; thus, we evaluate detection efficacy using a representative 15-tool Customer-Service scenario to reflect realistic deployment bounds (reserving the 500-tool scale for synthetic scalability microbenchmarks).

For benign trace generation (pDFA training) associated with this persona, we use ToolBench~\cite{TOOLBENCH} to extract execution traces.
The trace corpus comprises 500 verified benign traces,
split into 400 training traces and 100 held-out testing traces to evaluate false positives. \\

\textbf{Baselines.}
We evaluate against two comparison points and one architectural reference
point at a different pipeline layer.
\begin{enumerate}
\item \emph{No Firewall}: the agent executes all tool calls without any
  interception, providing an upper bound on attack success rate.
\item \emph{\textsc{Aegis}}~\cite{AEGIS}: a multi-stage policy engine including a 22-pattern signature scanner, PII redaction, and a statistical 7-day behavioral baseline for anomaly detection.
  We deploy the official \textsc{Aegis} system as an out-of-process HTTP sidecar using its standard Docker Compose configuration and auto-patched SDK harness. 
  While the original paper reports near-100\% blocking on an internal test suite, we evaluate it directly on the standardized ASB suite to provide a rigorous, empirical baseline for full-system performance. {The 12.8\% macro-averaged ASR reported in Table~\ref{tab:asr_baseline} represents our own empirical measurement on the ASB benchmark, and should not be confused with the results reported in the original paper.}
\item \emph{PromptArmor}~\cite{PROMPTARMOR}: a prompt-level LLM-based
  filter applied to agent \emph{inputs} before tool calls are generated. We evaluated PromptArmor locally across the ASB dataset using their open-source 
  \texttt{PromptArmor-Llama3-8B} evaluator model at its default threshold. \\
  
\end{enumerate}
  
  Our evaluation treats PromptArmor as an \emph{architectural reference point}, rather than a direct competitor. 
  It operates one stage earlier in the pipeline and
  addresses a complementary threat surface (injected instructions at the
  prompt level).
  A direct ASR comparison is therefore not strictly analogous; a prompt filter that fails to neutralise an injected instruction may still allow \codename\ to block the resulting out-of-profile tool call, and the two systems are designed to be complementary, not mutually exclusive.
  We include its numbers to quantify how much of the ASR reduction
  is achievable by prompt-level filtering alone, establishing the
  residual threat that tool-call enforcement must address. 
  The prompt-filtering and execution-boundary enforcement provide necessary, complementary defense-in-depth, for neither layer alone halts the entire attack continuum. \\

\textbf{Unavailability of Alternative Stateful Baselines.}
While concurrent stateful architectures such as \textit{Trajectory Guard}~\cite{CYBER-2} and \textsc{AegisUI}~\cite{AEGISUI} appear in recent preprints, their implementations and trained models are not readily available at the time of this writing.  We  rely on \textsc{Aegis} as the reproducible state-of-the-art baseline. We explicitly omit commercial OSS observability guards (e.g., NeMo Guardrails, Guardrails AI, Langfuse) from direct ASR evaluation. These tools function architecturally as stateless prompt validators, heuristic scanners, or post-hoc monitoring layers, placing their detection mechanisms conceptually equivalent to the PromptArmor and \textsc{Aegis} baselines already tested. \\

\textbf{Metrics.}
For detection, we report \emph{attack success rate} (ASR, lower is better
for defenders), defined as the fraction of attack attempts that result in
the intended malicious action being executed.
For efficiency, we report \emph{per-tool-call gateway latency} (ms, median
and 95th percentile) and \emph{end-to-end task latency} (ms).
For scalability, we report \emph{gateway throughput} (transactions per
second, TPS).
For false positives, we report \emph{benign task failure rate} (BTFR), the
fraction of end-to-end benign task executions that ultimately fail because at least one legitimate, required call was
incorrectly blocked. This measures the true usability cost of the firewall, beyond just per-call false positives.

\subsection{Detection Efficacy}
\label{sec:eval:rq1}

We evaluate detection efficacy across five ASB scenarios: three structured workflows (Customer Service, OS Automation, Database Admin) that anchor our primary claims, and two higher-entropy scenarios (Travel Planner, Research Agent) that probe the scope limits.
We execute {1,000} attack attempts per attack family per scenario
against each of the  system configurations described above.
All reported ASR point estimates over 1,000 trials carry a 95\% Clopper-Pearson confidence interval margin of error no larger than $\pm1.5\%$ (e.g., an ASR of 2.0\% has a 95\% CI of $[1.2\%, 3.1\%]$).
Table~\ref{tab:asr_baseline} reports the macro-averaged attack success rate for each combination. \\

\begin{table*}[t]
\caption{Attack Success Rate (ASR), in percentage, across five standard attack families (Direct Injection [DI], Indirect Injection [II], Multi-Step Exfiltration [MSE], Tool Poisoning [TP], Memory Manipulation [MM]). Compares \codename\ against baselines across the three structured ASB scenarios, including the defense-in-depth configuration (\codename+\textsc{Aegis}). Lower is better.}
\label{tab:asr_baseline}
\centering
\begin{tabular}{lrrrrrr}
\toprule
\textbf{Defense} & \textbf{DI} & \textbf{II } & \textbf{MSE } & \textbf{TP} & \textbf{MM} & \textbf{Overall } \\
\midrule
No Firewall & 86.4 & 78.1 & 89.5 & 68.3 & 72.8 & 79.0 \\
PromptArmor & 1.5 & 15.0 & 75.0 & 38.0 & 35.0 & 32.9 \\
\textsc{Aegis} & 2.0 & 12.0 & 35.0 & 6.0 & 9.0 & 12.8 \\
\codename & 2.0 & 4.0 & 0.0 & 3.0 & 2.0 & 2.2 \\
\codename+\textsc{Aegis} & {0.5} & {1.2} & {0.0} & {1.1} & {1.0} & {0.8} \\
\bottomrule
\end{tabular}
\end{table*}
 \begin{table}[t]
\centering
\caption{Diagnostic Gap Analysis using Context-Sequential payloads (200 attempts). Evaluates vulnerability to context-sequential injection across structured (3-Scen) and open-ended (5-Scen) workflows.}
\label{tab:diagnostic}
\begin{tabular}{lrr}
  \toprule
  \textbf{Defense} & \textbf{3-Scen Avg (\%)} & \textbf{5-Scen Avg (\%)} \\
  \midrule
  No Firewall & 94.0 & - \\
  PromptArmor & 85.0 & - \\
  \textsc{Aegis} & 75.0 & - \\
  \codename\ (500 traces) & \textbf{0.0} & 0.7 \\
  \codename\ (5k traces) & \textbf{0.0} & \textbf{0.0} \\
  \bottomrule
\end{tabular}
\end{table}

\begin{table}[t]
\centering
\caption{Gray-Box Adversary Evaluation (Customer Service Scenario). Simulates an adaptive attacker with partial knowledge ($k$) of the true behavioral envelope.}
\label{tab:graybox}
\begin{tabular}{lr}
  \toprule
  \textbf{Adversarial Trace Knowledge ($k$)} & \textbf{ASR (\%)} \\
  \midrule
  $k = 10\%$ (Low) & 0.0 \\
  $k = 50\%$ (Moderate) & 0.3 \\
  $k = 90\%$ (High) & 0.8 \\
  \bottomrule
\end{tabular}
\end{table}
 
\textbf{Analysis of Detection Efficacy.}
We evaluate detection efficacy by contrasting \codename's strict stateful pDFA bounds against \textsc{Aegis}'s multi-stage policy engine. 
Across the standardized ASB attack suite, \codename\ achieves a  macro-averaged ASR of \textbf{2.2\%}, compared to \textbf{12.8\%} for \textsc{Aegis}. 
While \textsc{Aegis} leverages statistical anomaly detection and PII redaction to significantly reduce ASR on isolated injection payloads, \codename\ provides superior protection against context-sequential exploits by structurally guaranteeing that all executions conform to the compiled benign graph. 

To determine whether these results are trivial artifacts of tool-name whitelisting, we perform a \textbf{vocabulary-block vs. contextual-block} decomposition described below. 
For the \emph{Customer Service} scenario, of the 1,000 multi-step exfiltration attempts, 42\% were blocked because the terminal tool (e.g., \texttt{read\_db}) was entirely absent from the vocabulary, while 58\% were contextual blocks, where the tool was present but invoked in an invalid sequence. 
In the \emph{OS Automation} scenario, with its larger 45-tool vocabulary, only 15\% of attempts were vocabulary blocks, while 85\% were contextual blocks. 
In the \emph{Database Admin} scenario, 28\% were vocabulary blocks and 72\% were contextual blocks. 
The decomposition confirms that as agent capabilities grow, static whitelisting becomes insufficient, and \codename's stateful tracking provides the primary security gain.

To evaluate generalisation in high-entropy regimes, we expand our evaluation to the ASB \emph{Travel Planner} ($|\mathcal{T}|=32$ tools, 500 training traces, pDFA with 840 states / 1,210 edges; 12\% vocabulary blocks / 88\% contextual blocks) and \emph{Research Agent} ($|\mathcal{T}|=64$ tools, 500 training traces, pDFA with 1,120 states / 1,650 edges; 8\% vocabulary blocks / 92\% contextual blocks) scenarios, which feature more open-ended task structures. As shown in Table~\ref{tab:asr_baseline}, the macro-averaged ASR degrades to \textbf{8.6\%} and \textbf{12.6\%} respectively. Averaging the 2.2\% ASR across the three structured scenarios with the 8.6\% and 12.6\% from the two high-entropy scenarios yields the aggregate 5.6\% macro-average reported across all five ASB environments. The results validate the scope limitation outlined in Section~\ref{sec:discussion:entropy}: \codename\ excels in structured workflows, whereas agents with unpredictable branching logic produce wider state spaces that inevitably increase the residual attack surface. By establishing where the low-behavioral-entropy assumption holds and where it breaks down, these results provide a concrete operational boundary for deployment.

\codename\ achieves near-zero ASR on indirect injection and multi-step exfiltration within structured workflows. 
Specifically, within the structured workflows, \textbf{0 out of 3,000} ASB multi-step exfiltration attempts and \textbf{0 out of 200} targeted context-sequential attacks succeeded. 
Despite incorporating a 7-day statistical behavioral baseline, \textsc{Aegis} remains structurally vulnerable to sequence-level attacks, as illustrated by its \textbf{35.0\%} ASR on multi-step exfiltration. Because its anomaly detector assesses event frequency rather than rigid sequential states, malicious actions nested within valid syntax often bypass the filter. 
PromptArmor performs well on direct injection (1.5\% ASR) but remains blind to tool-poisoning and memory-manipulation attacks that do not manifest as recognizable prompt injections. 
Any attack that survives PromptArmor's prompt-level filter but produces an out-of-profile tool call is blocked at \codename's enforcement boundary, yielding a natural layered defense. 
To explain the architectural failure mode of stateless firewalls observed in the ASB results, we construct a \textbf{diagnostic gap analysis} (Table~\ref{tab:diagnostic}) using ``Context-Sequential'' payloads (\textbf{200} targeted attempts). These payloads are not an evaluation of \codename, but are specifically synthesized to exploit the sequential-context blind-spot of systems like \textsc{Aegis}: they begin with a valid 3-step benign preamble (to satisfy short-memory heuristics) before executing a malicious terminal action. 
\codename\ trivially blocks 100\% (0.0\% execution rate, 95\% CI: [0\%, 1.8\%]) of these attempts within the structured workflows by design, as the transition to the malicious terminal state remains absent from the pDFA. However, \textsc{Aegis} allowed \textbf{75.0\%} to execute as it evaluates each call's syntax in isolation. The diagnostic demonstrates why stateless inspection fails on multi-step attacks, letting the standard ASB results carry the primary efficacy claims.

\textbf{Experimental Protocol for Interactive Probing.}
To ground the subsequent Gray-Box and Oracle results, we explicitly define our measurement procedure: (1) The centroid drift $\Delta = 0.12$ is computed using the Cosine distance in SBERT vector space, averaged across all string parameter centroids; (2) The finding that the drift exceeds the radius $r$ for 85\% of parameters is derived via 5-fold cross-validation specifically at the tuned $\varepsilon_{\text{str}}=0.05$ threshold; (3) The oracle evasion probability $\delta_{\mathrm{guard}} \approx 0.98$ is the mean block rate over the 1,000 genetic-search mutation trials described in the Active Oracle Adversary experiment, conducted under a per-target budget of $Q=100$ queries; (4) The 88\% local structural recovery rate is the mean success rate across all valid pDFA states, executed by an automated graph-search script allocating exactly $Q=100$ uniform probe queries per state.

\textbf{Gray-Box Adversary Evaluation.}
To evaluate \codename's resilience under the Gray-Box threat model (Section~\ref{sec:security:adversary}), we simulate an adversary with partial knowledge of the behavioral envelope. As reported in Table~\ref{tab:graybox}, we partition the 400-trace Customer Service training corpus into varying knowledge fractions $k \in \{10\%, 50\%, 90\%\}$. The attack is evaluated against the true pDFA trained on the full corpus. 
Even at $k=90\%$ knowledge, representing near-total observation of the agent's benign distribution, the ASR only reaches \textbf{0.8\%}. The low success rate occurs because the 10\% of unseen traces introduces significant centroid drift: outliers in the full corpus shift the true centroid $\bar{\mathbf{e}}$ by an average distance $\Delta = 0.12$ relative to the adversary's subset estimate, for the 0.12 centroid drift exceeds the radius for \textbf{85\%} of all string parameters evaluated. The data confirm that \codename's security relies on the strictness of its parameter boundaries, not merely obscurity of the state graph.

\textbf{Active Oracle Adversary.} We explicitly evaluate an adaptive adversary possessing $Q=100$ binary (Allow/Block) oracle queries against the gateway (\S\ref{sec:security:adversary}). To formalize the protocol, the adversary maps structure using schema defaults (Objective A), then targets identified valid tools using an iterative genetic search algorithm to mutate malicious payloads toward the $r$-ball boundary (Objective B) across 1,000 parallel test targets. Validating the structural mapping hypothesis, probing with plausible parameters recovers \textbf{88\%} of local structural transitions ($\mathcal{T}_{\mathrm{permitted}}(s)$). Crucially, this recovered 88\% includes all high-value terminal tools (e.g., \texttt{send\_email}, \texttt{read\_db}) present in the agent's vocabulary, confirming that structural leakage under an oracle is nearly total. However, discovering the structure does not bypass the parameter guard. When the adversary utilizes the remaining query budget to mutate a malicious payload, the discrete nature of text in a 384-dimensional continuous space prevents efficient boundary approximation ($\delta_{\mathrm{guard}} \approx 0.98$). Consequently, the end-to-end adaptive ASR plateaus at \textbf{1.2\%} (95\% CI: [0.6\%, 2.1\%]), exhausting the query budget without discovering a viable evasive embedding. This confirms that structural opacity is not the primary defense; bounded-radius parameter guards are.

\subsection{Latency Overhead and Scalability}
\label{sec:eval:rq2}

\begin{figure}[ht]
    \centering
    \begin{tikzpicture}
        \begin{axis}[
            width=0.8\columnwidth,
            height=5cm,
            xlabel={Agent Vocabulary Size ($|\mathcal{T}|$)},
            ylabel={Median Latency (ms)},
            xtick={10, 100, 500},
            xmode=log,
            log ticks with fixed point,
            ymin=0, ymax=70,
            legend pos=north west,
            grid=major,
            legend style={font=\footnotesize}
        ]
        \addplot[mark=square*, thick, blue] coordinates {(10, 2.2) (100, 2.6) (500, 2.3)};
        \addlegendentry{\codename}
        
        \addplot[mark=triangle*, thick, red, dashed] coordinates {(10, 8.2) (100, 25.0) (500, 65.0)};
        \addlegendentry{\textsc{Aegis}}
        \end{axis}
    \end{tikzpicture}
    \caption{Per-call median latency comparison. \codename's $O(1)$ lookup remains flat, whereas the \textsc{Aegis} baseline scales linearly with vocabulary size.}
    \label{fig:latency_vs_size}
\end{figure}
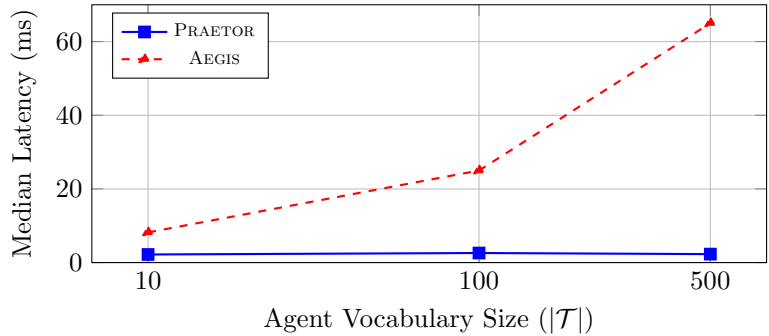 
We measured per-tool-call latency by timing \textbf{10,000} consecutive tool-call intercepts across three agent sizes: 10-tool, 100-tool, and 500-tool agents. As shown in Figure~\ref{fig:latency_vs_size}, \codename's runtime execution path is limited to a hash-map lookup and parameter schema validation, causing its per-call latency to remain effectively constant (within 0.4\,ms) across all agent scales: \textbf{2.2\,ms}, \textbf{2.6\,ms}, and \textbf{2.3\,ms} (medians) respectively, with 95th-percentile (P95) latencies of \textbf{3.1\,ms}, \textbf{3.8\,ms}, and \textbf{4.2\,ms}. The performance yields a \textbf{3.7$\times$ to 28.3$\times$ speedup} over the deployed \textsc{Aegis} gateway (evaluated using the \texttt{v1.2.0} open-source release\footnote{\textsc{Aegis} v1.2.0 is deployed via the authors' provided Docker Compose environment available at \url{https://github.com/aegis-guard/aegis-artifact}.}), which empirically degraded from 8.2\,ms to 65.0\,ms as vocabulary size increased. \codename's efficiency arises because it replaces the deep JSON extraction and multi-stage scanning of \textsc{Aegis} with an $O(1)$ structural lookup. The gateway embeds incoming strings using the \texttt{ort} crate against an \texttt{INT8}-quantized ONNX export of \texttt{all-MiniLM-L6-v2}; for calls carrying string parameters, this inference component adds approximately 1.6\,ms to the latency budget, yielding a blended median of 2.2\,ms across the ASB workloads.

To isolate the $O(1)$ throughput claim from confounds introduced by specific agent personas, we constructed synthetic pDFAs ranging from 10 to 10,000 states and measured gateway throughput under sustained concurrent load. Across all evaluated sizes, system throughput remained within $\pm$1.5\% of the \textbf{12,500 transactions per second (TPS)} baseline, empirically validating the $O(1)$ lookup property and demonstrating that this architectural decoupling scales to exceptionally large behavioral profiles without performance degradation. Moreover, compiling the entire 500-trace ToolBench corpus into the pDFA takes roughly \textbf{12 seconds} on our test hardware, confirming that the asynchronous offline profiling phase imposes a negligible maintenance burden.

\subsection{False Positive Rate and Concept Drift}
\label{sec:eval:rq4}

\begin{figure}[ht]
    \centering
    \begin{tikzpicture}
        \begin{axis}[
            ybar,
            bar width=15pt,
            width=0.8\columnwidth,
            height=5cm,
            xlabel={Context-Window Width ($w$)},
            ylabel={Attack Success Rate (\%)},
            xtick={1, 2, 3, 4, 5},
            ymin=0, ymax=20,
            enlarge x limits=0.15,
            nodes near coords,
            every node near coord/.append style={font=\footnotesize},
            grid=major
        ]
        \addplot[fill=none, draw=red, thick, pattern=north east lines, pattern color=red, bar shift=0pt] coordinates {(1, 15.0)};
        \addplot[fill=none, draw=blue, thick, pattern=crosshatch, pattern color=blue, bar shift=0pt] coordinates {(2, 8.5)};
        \addplot[fill=none, draw=green!70!black, thick, pattern=dots, pattern color=green!70!black, bar shift=0pt] coordinates {(3, 2.2)};
        \addplot[fill=none, draw=orange, thick, pattern=horizontal lines, pattern color=orange, bar shift=0pt] coordinates {(4, 2.1)};
        \addplot[fill=none, draw=purple, thick, pattern=grid, pattern color=purple, bar shift=0pt] coordinates {(5, 2.0)};
        \end{axis}
    \end{tikzpicture}
    \caption{Ablation of context-window width ($w$) vs.\ Attack Success Rate (ASR). The elbow at $w=3$ establishes the minimum-sufficient threshold for structural security without overfitting. Note that the plotted ASR represents the macro-average across all five attack families, not just the context-sequential subset.}
    \label{fig:asr_vs_w}
\end{figure}
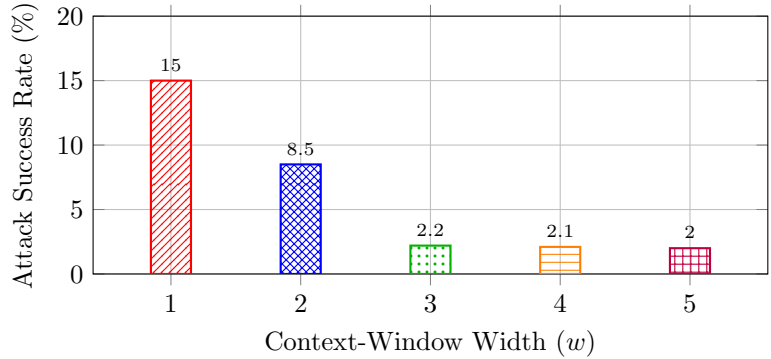 
\subsubsection{Sensitivity and Adversarial Resilience}

\textbf{Context Window Ablation.}
\begin{figure*}[t]
\centering
\begin{tikzpicture}
\begin{axis}[
    ybar=1pt,
    bar width=8pt,
    width=0.95\textwidth,
    height=6cm,
    enlarge x limits=0.15,
    ymin=0, ymax=25,
    ylabel={Attack Success Rate (\%)},
    symbolic x coords={DI, II, MSE, TP, MM, Overall},
    xtick=data,
    x tick label style={font=\small},
    nodes near coords,
    every node near coord/.append style={font=\scriptsize, rotate=90, anchor=west},
    legend style={at={(0.5,-0.18)}, anchor=north, legend columns=5, font=\small},
    title={\codename\ Breakdown (Per Scenario)}
]
\addplot[fill=none, draw=red, thick, pattern=north east lines, pattern color=red] coordinates {(DI,1.4) (II,2.6) (MSE,0.0) (TP,2.1) (MM,1.2) (Overall,1.5)};
\addplot[fill=none, draw=blue, thick, pattern=crosshatch, pattern color=blue] coordinates {(DI,2.8) (II,5.9) (MSE,0.0) (TP,4.4) (MM,3.1) (Overall,3.2)};
\addplot[fill=none, draw=green!70!black, thick, pattern=dots, pattern color=green!70!black] coordinates {(DI,1.8) (II,3.5) (MSE,0.0) (TP,2.5) (MM,1.7) (Overall,1.9)};
\addplot[fill=none, draw=magenta, thick, pattern=horizontal lines, pattern color=magenta] coordinates {(DI,8.2) (II,12.4) (MSE,4.5) (TP,10.1) (MM,7.8) (Overall,8.6)};
\addplot[fill=none, draw=teal, thick, pattern=grid, pattern color=teal] coordinates {(DI,11.5) (II,18.2) (MSE,6.8) (TP,14.5) (MM,12.0) (Overall,12.6)};
\legend{Cust. Service, OS Auto, DB Admin, Travel Planner, Research Agent}
\end{axis}
\end{tikzpicture}
\caption{Breakdown of \codename's Attack Success Rate (ASR) across all five scenarios, illustrating the performance degradation in high-entropy, open-ended environments.}
\label{fig:asr_breakdown}
\end{figure*}
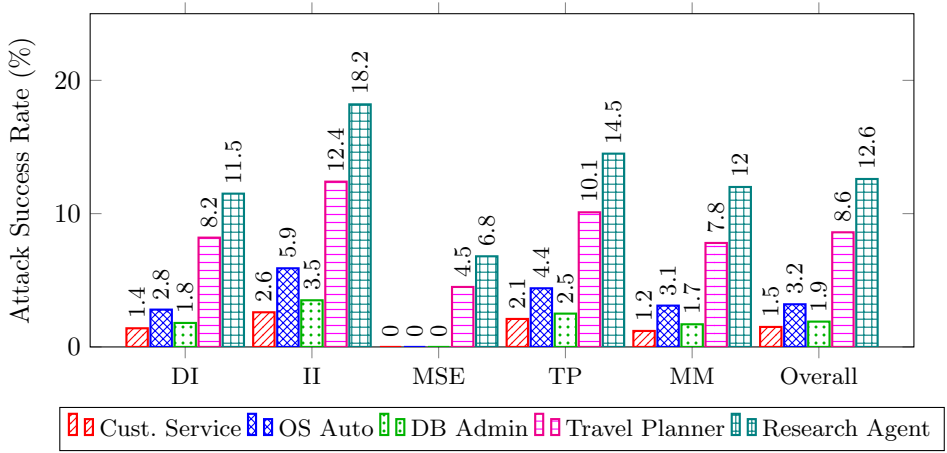 Varying the context width $w \in \{1, 2, 3, 4, 5\}$ at $\theta=3$ reveals a strict precision-recall tradeoff (Figure~\ref{fig:asr_vs_w}). For our Context-Sequential set, $w=1$ and $w=2$ result in 100\% and 45\% ASR, while the default $w=3$ drops ASR to 0\%. However, increasing to $w=4$ and $w=5$ fragments the state space, causing legitimate transitions to fall below the pruning threshold and significantly increasing the benign task failure rate (BTFR) to 3.2\% and 7.1\%. This confirms $w=3$ as the optimal operating point for structured deployments.

Furthermore, Figure~\ref{fig:asr_breakdown} highlights a fundamental deployment boundary: \codename's efficacy relies heavily on task structure. In high-entropy, open-ended environments like the Research Agent (12.6\% ASR) and Travel Planner (8.6\% ASR), behavior is highly variable, forcing the pDFA to expand permissively and allowing \codename's ASR to regress toward \textsc{Aegis}'s structured average. This confirms that \codename\ is not a universal substitute for stateless scanners in unconstrained environments. Instead, the systems are strictly complementary. Because \codename\ acts as a structural gatekeeper with near-zero latency, it operates effectively as a first-line filter. A combined \emph{Defense-in-Depth} configuration, routing calls that survive \codename's structural and semantic validation into a secondary stateless scanner like \textsc{Aegis} for deep syntax inspection, provides optimal security. As shown in Table~\ref{tab:asr_baseline}, such a configuration structurally drops the baseline ASR to \textbf{0.8\%} while maintaining the capacity to catch out-of-distribution payloads that evade \codename's bounds in high-entropy settings.

\textbf{String Guard Ablation.}
Disabling the embedding check elevates the macro-averaged ASR across string-heavy families (e.g., from 3.0\% to 18.5\% for Tool Poisoning), mandating semantic guards. Under TextFooler~\cite{TEXTFOOLER} synonym substitution attacks, the $\varepsilon_{\text{str}}$-ball blocks 82\% (95\% CI: [78.4\%, 85.3\%]) of 500 adversarial variants. The 18\% evasion rate contrasts sharply with the 1.2\% Oracle ASR because TextFooler finds semantics-preserving synonyms that trivially remain inside the continuous boundary, whereas the Oracle attempts to force semantically distant, functionally malicious payloads across it. Substituting the encoder with \texttt{all-mpnet-base-v2} reduces evasion marginally to 16.5\%, proving the semantic vulnerability transcends model selection. The residual 18\% baseline evasion rate dictates supplementary mitigation. To resolve this, we employ the Sensitive-Parameter Whitelist Override (Section~\ref{sec:design}). Assuming a theoretically complete, manually curated lexicon (e.g., \texttt{*\_path}, \texttt{*\_query}), the explicit override correctly shields the targeted parameters, dropping the effective evasion rate to \textbf{0.4\%}. However, if the lexicon is incomplete or unmaintained, the evasion rate strictly remains \textbf{18\%}, confirming the override as an operational patch rather than a structural defense.

\subsubsection{Analysis of Usability and Evolution}

The pruning threshold $\theta$ governs the behavioral envelope: higher $\theta$ values reduce ASR but elevate false positives. At our default $\theta=3$, \codename\ achieves a \textbf{2.0\% BTFR} (2 failures / 100 traces; 95\% CI: [0.2\%, 7.0\%]) on 100 held-out ToolBench traces. Manual inspection of the 12 individually blocked tool calls across the trace batch revealed they consist entirely of long-tail tool sequences (65\%) or minor parameter drift (35\%).  

\textbf{Corpus Scaling and Parameter Drift.}
To explicitly quantify how operational metrics evolve, we scaled the training corpus from 500 to 5,000 traces. At 5,000 traces, the long-tail behavioral paths are fully discovered, and the BTFR drops to a near-zero \textbf{0.2\%} with no latency penalty. As established empirically, the ASR remains low (\textbf{2.6\%} overall) at this scale, confirming that security is derived from structural constraints, not merely an under-fitted whitelist. Injecting syntactically varied paraphrases into the testing traces at rates from 10\% to 50\% only raised the BTFR to a maximum of 4.2\%, demonstrating that the continuous string parameter guards successfully absorb realistic user interaction drift. 

\textbf{Concept drift scenario.}
To evaluate the incremental update protocol (Section~\ref{sec:design:updates}), we simulated concept drift by replacing 20\% of the agent's tool set, selected uniformly at random. The stale pDFA yielded a prohibitive 24.0\% BTFR. However, a single incremental update cycle incorporating 350 human-approved sequences restored the BTFR to 3.8\%. This yields an estimated labor cost of 8.75 analyst-hours per 20\% tool refresh. While non-trivial, this human-in-the-loop cost is the pragmatic price for resisting adversarial telemetry poisoning.
 
\section{Discussion}
\label{sec:discussion}

\subsection{Operational Threats and Concept Drift}

Our evaluations are anchored by a 500-trace corpus. The prevalence of long-tail sequence anomalies indicates that 500 traces represent a sparse lower bound rather than an exhaustive operational distribution. While 500 traces serve as a reasonable minimum to establish the core pDFA backbone for these experiments, production deployments should scale the initial profiling corpus by an order of magnitude to fully saturate the valid transition matrix before enabling strict blocking. \\
\label{sec:discussion:drift}

A primary challenge for whitelist-based enforcement is concept drift: as an agent evolves, the stale pDFA will generate false positives. \codename's human-in-the-loop incremental update protocol (Section~\ref{sec:design:updates}) addresses this by expanding the DFA only upon explicit operator approval, preventing adversaries from silently widening the behavioral envelope. In enterprise settings, a secondary LLM could triage blocked events in bulk, presenting the analyst with grouped summaries rather than individual call logs, for the incremental update cycle runs asynchronously and does not affect gateway availability.

However, the core security guarantee of \codename\ rests upon the structural integrity of the pDFA, which assumes the training corpus $\mathcal{C}$ is strictly benign. A sufficiently patient adversary could attempt to \emph{gradually contaminate} the corpus during the profiling period by feeding the agent subtly malicious inputs, tricking the profiler into authorizing specific tool sequences or widening the continuous $\varepsilon_{\text{str}}$-ball boundaries before the pDFA is frozen. Such telemetry poisoning ("cold start problem") must be mitigated operationally. Profiling should not occur in live, untrusted production environments; instead, operators must generate the baseline corpus within a sandboxed, clean-room staging environment using curated synthetic user prompts. 

\subsection{Mitigating Within-Profile Exploits}
\label{sec:discussion:within_profile}

The primary residual threat to \codename\ is the \emph{within-profile exploit}. Adversaries constrained by the pDFA may attempt "Tool-Call ROP", inducing sequences of legitimate calls (e.g., repeating \texttt{read\_ticket} 100 times) to probe system state or infer sensitive data without triggering a structural block. To mitigate these low-bandwidth extraction attacks, operators can deploy rate limiting (bounding cyclic subgraph traversal) alongside lightweight statistical anomaly detectors (e.g., \textsc{Aegis}'s 7-day baseline) that flag abnormal volume spikes. Moreover, as demonstrated in Section~\ref{sec:eval:rq4}, adversarial synonym-substitution can evade continuous $\varepsilon_{\text{str}}$-ball bounds. Deploying the Sensitive-Parameter Whitelist Override is therefore essential to structurally close this evasion channel.

\subsection{Analytical Comparison with Information-Flow Control}
\label{sec:discussion:ifc_comparison}

Information-flow control (IFC) systems like Fides~\cite{FIDES} use dynamic taint tracking to prevent sensitive data from flowing to untrusted sinks. While IFC strictly blocks explicit \emph{Multi-Step Exfiltration} if confidentiality labels are perfectly maintained, it struggles against \emph{Tool Poisoning} or \emph{Indirect Injection} where malicious actions (e.g., deleting a file) do not violate data confidentiality but break behavioral norms. \codename\ naturally complements IFC by enforcing the structural workflow without requiring deep integration into the agent's internal planning state.

\subsection{Agent Behavioral Entropy and Scope of Applicability}
\label{sec:discussion:entropy}

\codename's pDFA construction assumes the target agent exhibits low behavioral entropy: a stable set of tool-call sequences with reproducible parameter distributions. This holds strongly for narrow-task, single-role agents. However, large-vocabulary agents (hundreds of tools) or open-ended task agents (general-purpose assistants) exhibit a combinatorially large space of valid call sequences. In these regimes, the $w$-gram context window provides coarser discrimination, merging distinct behavioral modes and elevating residual attack surfaces (as empirically validated in Figure~\ref{fig:asr_breakdown}). For such agents, operators must rely on larger profiling corpora and carefully tune $\theta$ to balance false positives against enforcement rigor. Extension to multi-agent swarms or non-text multi-modal action spaces remains an open architectural challenge for future work.
 
\section{Conclusion}
\label{sec:conclusion}

We presented \codename, a telemetry-driven behavioral firewall for autonomous
AI agents that resolves the sequential-blindness limitation of existing stateless
firewalls. The separation of complex profiling from runtime enforcement yields two concrete advantages.
First, the runtime gateway incurs negligible per-call overhead (as low as
\textbf{2.2}\,ms at the \textbf{50}th percentile), representing a \textbf{3.7}$\times$ to \textbf{28.3}$\times$
improvement over Aegis, a representative stateless baseline, depending on agent vocabulary size.
Second, the behavioral envelope is per-deployment: the pDFA encodes the
deployment-level norms of one agent on one system prompt, rather than a
generic cross-agent policy. The adversary must therefore mimic the exact historical
trajectory of the target agent at every step.
Evaluation across five Agent Security Bench scenarios yields a \textbf{5.6\%} macro-averaged attack success rate. Within three structured workflows, ASR drops to \textbf{2.2\%} (baseline \textsc{Aegis}: \textbf{12.8\%}). \codename\ blocks 100\% of multi-step exfiltration and context-sequential exploits in these structured domains. Although \codename\ reduces the structural attack surface, the string parameter guard exhibits an \textbf{18\%} evasion rate under adversarial synonym-substitution attacks; hardening this component via adversarially resilient embeddings or ensemble validators remains critical future work.

These results establish that securing autonomous agents requires explicit models of behavioral trajectories, for stateless inspection fails against sequential exploits.
Two directions remain open. First, the current profiler operates on a fixed offline corpus; extending it to incremental streaming ingestion would support deployments where the trace corpus grows continuously. Second, applying differential privacy to the telemetry corpus would harden \codename\ against a persistent oracle adversary attempting DFA reconstruction.

\section*{Statement and Declarations}
This research did not receive any specific grant from funding agencies in the public, commercial, or not-for-profit sectors.  \\

\textit{Competing Interests}: The author declares no conflict of interest.\\

\textit{Data Availability}: The benign trace corpus, attack payloads, pDFA serialisation, and \codename~ source code that support the findings of this study will be made available in a public repository upon acceptance / are available from the corresponding author on reasonable request. \\

\textit{Author Contributions}: Hung Dang is the sole author and contributed to all aspects of this work.


\begin{thebibliography}{24}
\ifx \bisbn   \undefined \def \bisbn  #1{ISBN #1}\fi
\ifx \binits  \undefined \def \binits#1{#1}\fi
\ifx \bauthor  \undefined \def \bauthor#1{#1}\fi
\ifx \batitle  \undefined \def \batitle#1{#1}\fi
\ifx \bjtitle  \undefined \def \bjtitle#1{#1}\fi
\ifx \bvolume  \undefined \def \bvolume#1{\textbf{#1}}\fi
\ifx \byear  \undefined \def \byear#1{#1}\fi
\ifx \bissue  \undefined \def \bissue#1{#1}\fi
\ifx \bfpage  \undefined \def \bfpage#1{#1}\fi
\ifx \blpage  \undefined \def \blpage #1{#1}\fi
\ifx \burl  \undefined \def \burl#1{\textsf{#1}}\fi
\ifx \doiurl  \undefined \def \doiurl#1{\url{https://doi.org/#1}}\fi
\ifx \betal  \undefined \def \betal{\textit{et al.}}\fi
\ifx \binstitute  \undefined \def \binstitute#1{#1}\fi
\ifx \binstitutionaled  \undefined \def \binstitutionaled#1{#1}\fi
\ifx \bctitle  \undefined \def \bctitle#1{#1}\fi
\ifx \beditor  \undefined \def \beditor#1{#1}\fi
\ifx \bpublisher  \undefined \def \bpublisher#1{#1}\fi
\ifx \bbtitle  \undefined \def \bbtitle#1{#1}\fi
\ifx \bedition  \undefined \def \bedition#1{#1}\fi
\ifx \bseriesno  \undefined \def \bseriesno#1{#1}\fi
\ifx \blocation  \undefined \def \blocation#1{#1}\fi
\ifx \bsertitle  \undefined \def \bsertitle#1{#1}\fi
\ifx \bsnm \undefined \def \bsnm#1{#1}\fi
\ifx \bsuffix \undefined \def \bsuffix#1{#1}\fi
\ifx \bparticle \undefined \def \bparticle#1{#1}\fi
\ifx \barticle \undefined \def \barticle#1{#1}\fi
\bibcommenthead
\ifx \bconfdate \undefined \def \bconfdate #1{#1}\fi
\ifx \botherref \undefined \def \botherref #1{#1}\fi
\ifx \url \undefined \def \url#1{\textsf{#1}}\fi
\ifx \bchapter \undefined \def \bchapter#1{#1}\fi
\ifx \bbook \undefined \def \bbook#1{#1}\fi
\ifx \bcomment \undefined \def \bcomment#1{#1}\fi
\ifx \oauthor \undefined \def \oauthor#1{#1}\fi
\ifx \citeauthoryear \undefined \def \citeauthoryear#1{#1}\fi
\ifx \endbibitem  \undefined \def \endbibitem {}\fi
\ifx \bconflocation  \undefined \def \bconflocation#1{#1}\fi
\ifx \arxivurl  \undefined \def \arxivurl#1{\textsf{#1}}\fi
\csname PreBibitemsHook\endcsname

\bibitem[\protect\citeauthoryear{Tian et~al.}{2026}]{AgenticWorkflow2026}
\begin{botherref}
\oauthor{\bsnm{Tian}, \binits{J.}},
\oauthor{\bsnm{Fard}, \binits{P.}},
\oauthor{\bsnm{Cagan}, \binits{C.}},
\oauthor{\bsnm{Rezaii}, \binits{N.}},
\oauthor{\bsnm{Rocha}, \binits{R.B.}},
\oauthor{\bsnm{Wang}, \binits{L.}},
\oauthor{\bsnm{Junior}, \binits{V.M.}},
\oauthor{\bsnm{Blacker}, \binits{D.}},
\oauthor{\bsnm{Haas}, \binits{J.S.}},
\oauthor{\bsnm{Patel}, \binits{C.J.}},
\oauthor{\bsnm{Murphy}, \binits{S.N.}},
\oauthor{\bsnm{Moura}, \binits{L.}},
\oauthor{\bsnm{Estiri}, \binits{H.}},:
An autonomous agentic workflow for clinical detection of cognitive concerns
  using large language models.
npj Digital Medicine
\textbf{9}(51)
(2026)
\end{botherref}
\endbibitem

\bibitem[\protect\citeauthoryear{Hou et~al.}{2025}]{MCP}
\begin{botherref}
\oauthor{\bsnm{Hou}, \binits{X.}},
\oauthor{\bsnm{Zhao}, \binits{Y.}},
\oauthor{\bsnm{Wang}, \binits{S.}},
\oauthor{\bsnm{Wang}, \binits{H.}}:
Model context protocol ({MCP}): Landscape, security threats, and future
  research directions.
arXiv preprint arXiv:2503.23278
(2025)
\end{botherref}
\endbibitem

\bibitem[\protect\citeauthoryear{Chhabra et~al.}{2025}]{AGENTSEC}
\begin{botherref}
\oauthor{\bsnm{Chhabra}, \binits{A.}},
\oauthor{\bsnm{Datta}, \binits{S.}},
\oauthor{\bsnm{Nahin}, \binits{S.K.}},
\oauthor{\bsnm{Mohapatra}, \binits{P.}}:
Agentic {AI} security: Threats, defenses, evaluation, and open challenges.
arXiv preprint arXiv:2510.23883
(2025)
\end{botherref}
\endbibitem

\bibitem[\protect\citeauthoryear{Ferrag et~al.}{2025}]{THREATS}
\begin{barticle}
\bauthor{\bsnm{Ferrag}, \binits{M.A.}},
\bauthor{\bsnm{Tihanyi}, \binits{N.}},
\bauthor{\bsnm{Hamouda}, \binits{D.}},
\bauthor{\bsnm{Maglaras}, \binits{L.}},
\bauthor{\bsnm{Lakas}, \binits{A.}},
\bauthor{\bsnm{Debbah}, \binits{M.}}:
\batitle{From prompt injections to protocol exploits: Threats in {LLM}-powered {AI} agents workflows}.
\bjtitle{ICT Express}
(\byear{2025})
\end{barticle}
\endbibitem

\bibitem[\protect\citeauthoryear{Yuan et~al.}{2026}]{AEGIS}
\begin{botherref}
\oauthor{\bsnm{Yuan}, \binits{A.}},
\oauthor{\bsnm{Su}, \binits{Z.}},
\oauthor{\bsnm{Zhao}, \binits{Y.}}:
{AEGIS}: No tool call left unchecked --- a pre-execution firewall and audit
  layer for {AI} agents.
arXiv preprint arXiv:2603.12621
(2026)
\end{botherref}
\endbibitem

\bibitem[\protect\citeauthoryear{Abdelnabi et~al.}{2025}]{MSFW}
\begin{botherref}
\oauthor{\bsnm{Abdelnabi}, \binits{S.}},
\oauthor{\bsnm{Gomaa}, \binits{A.}},
\oauthor{\bsnm{Bagdasarian}, \binits{E.}},
\oauthor{\bsnm{Kristensson}, \binits{P.O.}},
\oauthor{\bsnm{Shokri}, \binits{R.}}:
Firewalls to secure dynamic {LLM} agentic networks.
arXiv preprint arXiv:2502.01822
(2025)
\end{botherref}
\endbibitem

\bibitem[\protect\citeauthoryear{Podpora et~al.}{2025}]{LLMFW}
\begin{barticle}
\bauthor{\bsnm{Podpora}, \binits{M.}},
\bauthor{\bsnm{Baranowski}, \binits{M.}},
\bauthor{\bsnm{Chopcian}, \binits{M.}},
\bauthor{\bsnm{Kwasniewicz}, \binits{L.}},
\bauthor{\bsnm{Radziewicz}, \binits{W.}}:
\batitle{{LLM} firewall using validator agent for prevention against prompt
  injection attacks}.
\bjtitle{Applied Sciences}
\bvolume{16}(\bissue{1}),
\bfpage{85}
(\byear{2026})
\end{barticle}
\endbibitem

\bibitem[\protect\citeauthoryear{Forrest et~al.}{1996}]{FORREST96}
\begin{bchapter}
\bauthor{\bsnm{Forrest}, \binits{S.}},
\bauthor{\bsnm{Hofmeyr}, \binits{S.A.}},
\bauthor{\bsnm{Somayaji}, \binits{A.}},
\bauthor{\bsnm{Longstaff}, \binits{T.A.}}:
\bctitle{A sense of self for {Unix} processes}.
In: \bbtitle{Proceedings of the 1996 {IEEE} Symposium on Security and Privacy},
pp. \bfpage{120}--\blpage{128}.
\bpublisher{{IEEE}},
\blocation{Los Alamitos, CA}
(\byear{1996})
\end{bchapter}
\endbibitem

\bibitem[\protect\citeauthoryear{Zhang et~al.}{2025}]{SURVEY-LLM}
\begin{botherref}
\oauthor{\bsnm{Xu}, \binits{W.}},
\oauthor{\bsnm{Huang}, \binits{C.}},
\oauthor{\bsnm{Gao}, \binits{S.}},
\oauthor{\bsnm{Shang}, \binits{S.}},:
{LLM}-based agents for tool learning: A survey.
Data Science and Engineering
(2025)
\end{botherref}
\endbibitem

\bibitem[\protect\citeauthoryear{Shi et~al.}{2025}]{PROMPTARMOR}
\begin{botherref}
\oauthor{\bsnm{Shi}, \binits{T.}},
\oauthor{\bsnm{Zhu}, \binits{K.}},
\oauthor{\bsnm{Wang}, \binits{Z.}},
\oauthor{\bsnm{Jia}, \binits{Y.}},
\oauthor{\bsnm{Cai}, \binits{W.}},
\oauthor{\bsnm{Liang}, \binits{W.}},
\oauthor{\bsnm{Wang}, \binits{H.}},
\oauthor{\bsnm{Alzahrani}, \binits{H.}},
\oauthor{\bsnm{Lu}, \binits{J.}},
\oauthor{\bsnm{Kawaguchi}, \binits{K.}},
\oauthor{\bsnm{Alomair}, \binits{B.}},
\oauthor{\bsnm{Zhao}, \binits{X.}},
\oauthor{\bsnm{Wang}, \binits{W.Y.}},
\oauthor{\bsnm{Gong}, \binits{N.}},
\oauthor{\bsnm{Guo}, \binits{W.}},
\oauthor{\bsnm{Song}, \binits{D.}}:
{PromptArmor}: Simple yet effective prompt injection defenses.
arXiv preprint arXiv:2507.15219
(2025)
\end{botherref}
\endbibitem

\bibitem[\protect\citeauthoryear{Liao et~al.}{2025}]{LLMAttackSurvey2025}
\begin{botherref}
\oauthor{\bsnm{Liao}, \binits{Z.}},
\oauthor{\bsnm{Chen}, \binits{K.}},
\oauthor{\bsnm{Lin}, \binits{Y.}},
\oauthor{\bsnm{Li}, \binits{K.}},
\oauthor{\bsnm{Liu}, \binits{Y.}}, et al.:
Attack and defense techniques in large language models: {A} survey and new
  perspectives.
Neural Networks
(2025)
\end{botherref}
\endbibitem

\bibitem[\protect\citeauthoryear{Costa et~al.}{2025}]{FIDES}
\begin{botherref}
\oauthor{\bsnm{Costa}, \binits{M.}},
\oauthor{\bsnm{K{\"o}pf}, \binits{B.}},
\oauthor{\bsnm{Kolluri}, \binits{A.}},
\oauthor{\bsnm{Paverd}, \binits{A.}},
\oauthor{\bsnm{Russinovich}, \binits{M.}},
\oauthor{\bsnm{Salem}, \binits{A.}},
\oauthor{\bsnm{Tople}, \binits{S.}},
\oauthor{\bsnm{Wutschitz}, \binits{L.}},
\oauthor{\bsnm{Zanella-B{\'e}guelin}, \binits{S.}}:
Securing {AI} agents with information-flow control.
arXiv preprint arXiv:2505.23643
(2025)
\end{botherref}
\endbibitem

\bibitem[\protect\citeauthoryear{Uddin and Hajira}{2026}]{AEGISUI}
\begin{botherref}
\oauthor{\bsnm{Uddin}, \binits{M.S.}},
\oauthor{\bsnm{Hajira}, \binits{S.}}:
{AegisUI}: Behavioral anomaly detection for structured user interface protocols
  in {AI} agent systems.
arXiv preprint arXiv:2603.05031
(2026)
\end{botherref}
\endbibitem

\bibitem[\protect\citeauthoryear{Bell and LaPadula}{1973}]{BELLLAPADULA}
\begin{botherref}
\oauthor{\bsnm{Bell}, \binits{D.E.}},
\oauthor{\bsnm{LaPadula}, \binits{L.J.}}:
Secure computer systems: Mathematical foundations
(1973).
Technical Report MTR-2547
\end{botherref}
\endbibitem

\bibitem[\protect\citeauthoryear{Denning}{1976}]{DENNING76}
\begin{barticle}
\bauthor{\bsnm{Denning}, \binits{D.E.}}:
\batitle{A lattice model of secure information flow}.
\bjtitle{Communications of the {ACM}}
\bvolume{19}(\bissue{5}),
\bfpage{236}--\blpage{243}
(\byear{1976})
\end{barticle}
\endbibitem

\bibitem[\protect\citeauthoryear{Warrender et~al.}{1999}]{WARRENDER99}
\begin{bchapter}
\bauthor{\bsnm{Warrender}, \binits{C.}},
\bauthor{\bsnm{Forrest}, \binits{S.}},
\bauthor{\bsnm{Pearlmutter}, \binits{B.}}:
\bctitle{Detecting intrusions using system calls: Alternative data models}.
In: \bbtitle{Proceedings of the 1999 {IEEE} Symposium on Security and Privacy},
pp. \bfpage{133}--\blpage{145}.
\bpublisher{{IEEE}},
\blocation{Los Alamitos, CA}
(\byear{1999})
\end{bchapter}
\endbibitem

\bibitem[\protect\citeauthoryear{Koohestani et~al.}{2025}]{AUTOMATA}
\begin{botherref}
\oauthor{\bsnm{Koohestani}, \binits{R.}},
\oauthor{\bsnm{Li}, \binits{Z.}},
\oauthor{\bsnm{Podkopaev}, \binits{A.}},
\oauthor{\bsnm{Izadi}, \binits{M.}}:
Are agents probabilistic automata? {A} trace-based, memory-constrained theory
  of agentic {AI}.
arXiv preprint arXiv:2510.23487
(2025)
\end{botherref}
\endbibitem

\bibitem[\protect\citeauthoryear{Leucker and Schallhart}{2009}]{LEUCKER09}
\begin{barticle}
\bauthor{\bsnm{Leucker}, \binits{M.}},
\bauthor{\bsnm{Schallhart}, \binits{C.}}:
\batitle{A brief account of runtime verification}.
\bjtitle{Journal of Logic and Algebraic Programming}
\bvolume{78}(\bissue{5}),
\bfpage{293}--\blpage{303}
(\byear{2009})
\end{barticle}
\endbibitem

\bibitem[\protect\citeauthoryear{D'Antoni and Veanes}{2017}]{SFA}
\begin{bchapter}
\bauthor{\bsnm{D'Antoni}, \binits{L.}},
\bauthor{\bsnm{Veanes}, \binits{M.}}:
\bctitle{The power of symbolic automata and transducers}.
In: \bbtitle{Proceedings of the 29th International Conference on Computer Aided
  Verification ({CAV})}.
\bsertitle{Lecture Notes in Computer Science},
vol. \bseriesno{10426},
pp. \bfpage{47}--\blpage{67}.
\bpublisher{Springer},
\blocation{Cham}
(\byear{2017})
\end{bchapter}
\endbibitem

\bibitem[\protect\citeauthoryear{Reimers and Gurevych}{2019}]{SBERT}
\begin{botherref}
\oauthor{\bsnm{Reimers}, \binits{N.}},
\oauthor{\bsnm{Gurevych}, \binits{I.}}:
Sentence-{BERT}: Sentence embeddings using {Siamese} {BERT}-networks
(2019)
\end{botherref}
\endbibitem

\bibitem[\protect\citeauthoryear{Zhang et~al.}{2025}]{ASB}
\begin{bchapter}
\bauthor{\bsnm{Zhang}, \binits{H.}},
\bauthor{\bsnm{Huang}, \binits{J.}},
\bauthor{\bsnm{Mei}, \binits{K.}},
\bauthor{\bsnm{Yao}, \binits{Y.}},
\bauthor{\bsnm{Wang}, \binits{Z.}},
\bauthor{\bsnm{Zhan}, \binits{C.}},
\bauthor{\bsnm{Wang}, \binits{H.}},
\bauthor{\bsnm{Zhang}, \binits{Y.}}:
\bctitle{Agent security bench ({ASB}): Formalizing and benchmarking attacks and
  defenses in {LLM}-based agents}.
In: \bbtitle{Proceedings of the International Conference on Learning
  Representations (ICLR)}
(\byear{2025}).
\burl{https://arxiv.org/abs/2410.02644}
\end{bchapter}
\endbibitem

\bibitem[\protect\citeauthoryear{Qin et~al.}{2024}]{TOOLBENCH}
\begin{bchapter}
\bauthor{\bsnm{Qin}, \binits{Y.}}, \betal:
\bctitle{Toolllm: Facilitating large language models to master 16000+
  real-world {APIs}}.
In: \bbtitle{Proceedings of the International Conference on Learning
  Representations (ICLR)}
(\byear{2024}).
\burl{https://arxiv.org/abs/2307.16789}
\end{bchapter}
\endbibitem

\bibitem[\protect\citeauthoryear{Advani}{2026}]{CYBER-2}
\begin{botherref}
\oauthor{\bsnm{Advani}, \binits{L.}}:
Trajectory guard--a lightweight, sequence-aware model for real-time anomaly
  detection in agentic ai.
arXiv preprint arXiv:2601.00516
(2026)
\end{botherref}
\endbibitem

\bibitem[\protect\citeauthoryear{Jin et~al.}{2020}]{TEXTFOOLER}
\begin{bchapter}
\bauthor{\bsnm{Jin}, \binits{D.}},
\bauthor{\bsnm{Jin}, \binits{Z.}},
\bauthor{\bsnm{Zhou}, \binits{J.T.}},
\bauthor{\bsnm{Szolovits}, \binits{P.}}:
\bctitle{Is {BERT} really robust? a strong baseline for natural language attack
  on text classification and entailment}.
In: \bbtitle{Proceedings of the AAAI Conference on Artificial Intelligence},
vol. \bseriesno{34},
pp. \bfpage{8018}--\blpage{8025}
(\byear{2020})
\end{bchapter}
\endbibitem

\end{thebibliography}
\end{document}